\documentclass[11pt]{article}
\pdfoutput=1

\usepackage{jheppub}
\usepackage{amssymb,amsfonts,amsmath,amsthm,lineno}
\usepackage{enumerate}
\usepackage{mathrsfs}
\usepackage{bbold}
\usepackage{tikz}
\usetikzlibrary{shapes.geometric, arrows}
\usepackage{hyperref}
\usepackage{slashed}

\makeatletter
\newcommand{\Vast}{\bBigg@{4.75}}
\makeatother

\newcommand{\be}{\begin{equation}}
\newcommand{\ee}{\end{equation}}
\newcommand{\bea}{\begin{eqnarray}}
\newcommand{\eea}{\end{eqnarray}}

\newcommand{\CA}{\mathcal{A}}

\newcommand{\CC}{\mathcal{C}}
\newcommand{\CD}{\mathcal{D}}
\newcommand{\CE}{\mathcal{E}}

\newcommand{\CG}{\mathcal{G}}
\newcommand{\CH}{\mathcal{H}}

\newcommand{\CJ}{\mathcal{J}}
\newcommand{\CK}{\mathcal{K}}
\newcommand{\CL}{\mathcal{L}}
\newcommand{\CN}{\mathcal{N}}

\newcommand{\CO}{\mathcal{O}}
\newcommand{\CP}{\mathcal{P}}
\newcommand{\CQ}{\mathcal{Q}}

\newcommand{\CV}{\mathcal{V}}

\newcommand{\CY}{\mathcal{Y}}
\newcommand{\CZ}{\mathcal{Z}}

\newcommand{\lr}{\left (}
\newcommand{\rr}{\right )}
\newcommand{\ls}{\left [}
\newcommand{\rs}{\right ]}
\newcommand{\lc}{\left \{}
\newcommand{\rc}{\right \}}

\newcommand\qt\tau

\newcommand{\p}{\partial}

\renewcommand{\tilde}[1]{\widetilde{#1}}
\newcommand{\tr}{\text{tr}}

\makeatletter
\renewcommand{\@seccntformat}[1]{\csname the#1\endcsname.\,\,}
\makeatother
\allowdisplaybreaks

\let \savenumberline \numberline
\def \numberline#1{\savenumberline{#1.}}

\makeatletter
\def\@fpheader{\relax}
\makeatother

\def\bea{\begin{eqnarray}}
\def\eea{\end{eqnarray}}

\usetikzlibrary{shapes,arrows,chains}
\usetikzlibrary{decorations.markings}
\usetikzlibrary{decorations.pathmorphing}
\usetikzlibrary{shapes.multipart}
\tikzset{snake it/.style={decorate, decoration=snake}}

\title{\ \vspace{1.6cm} \\
\scalebox{0.95}{Renormalization of Supersymmetric Lifshitz Sigma Models}}
\author{Ziqi Yan}
\emailAdd{ziqi.yan@su.se}
\affiliation{
Nordita, KTH Royal Institute of Technology and Stockholm University\\
Hannes Alfv\'{e}ns v\"{a}g 12, SE-106 91 Stockholm, Sweden}
\abstract{We study the renormalization of an $\CN = 1$ supersymmetric Lifshitz sigma model in three dimensions. The sigma model exhibits worldvolume anisotropy in space and time around the high-energy $z=2$ Lifshitz point, such that the worldvolume is endowed with a foliation structure along a preferred time direction. In curved backgrounds, the target-space geometry is equipped with two distinct metrics, and the interacting sigma model is power-counting renormalizable. At low energies, the theory naturally flows toward the relativistic sigma model where Lorentz symmetry emerges. In the superspace formalism, we develop a heat kernel method that is covariantized with respect to the bimetric target-space geometry, using which we evaluate the one-loop beta-functions of the Lifshitz sigma model. This study forms an essential step toward a thorough understanding of the quantum critical supermembrane as a candidate high-energy completion of the relativistic supermembrane.} 

\begin{document}

\maketitle
\vfill\eject

\section{Introduction} \label{sec:intro}

The quantization of supermembranes in eleven-dimensional spacetime is key to a better understanding of M-theory, which provides a unification of different superstring theories in a nonperturbative manner. Supermembranes are described by a three-dimensional sigma model \cite{Bergshoeff:1987cm}, whose quantization is fundamentally distinct from quantizing strings. One of the major difficulties is that three-dimensional relativistic sigma models are not renormalizable, which invalidates any perturbative treatment. Nevertheless, 
it is possible to probe some features of the supermembrane in the light-cone quantization, at least when a matrix regularization is employed on the spatial manifold of the worldvolume \cite{Goldstone, Hoppe}. This regularization restricts the infinite-dimensional basis on the continuum manifold to a finite-dimensional Lie group, and the supermembrane is described as a subtle limit of supersymmetric matrix quantum mechanics \cite{deWit:1988wri}, closely related to D0-branes in the infinite momentum frame and the Matrix theory description of M-theory \cite{Banks:1996vh}.\,\footnote{Another way of obtaining Matrix theory is by considering the discrete light cone quantization (DLCQ) of M-theory, which is usually defined as a subtle limit of the compactification over a spatial circle \cite{Susskind:1997cw, Seiberg:1997ad, Sen:1997we, Hellerman:1997yu}. An alternative treatment of the DLCQ of string/M-theory as nonrelativistic string/M-theory is recently revived in \emph{e.g.} \cite{Bergshoeff:2018yvt, Ebert:2021mfu}.} However, a complete understanding of the continuous spectrum of Matrix theory is still lacking \cite{deWit:1988xki, Smilga:1989ew}, which requires a multiparticle interpretation and thus a second quantization of the membrane \cite{Banks:1996vh}. This light-cone treatment is also limited to special backgrounds, due to the absence of spacetime covariance. Finally, when the sigma model is coupled to a dynamical worldvolume described by three-dimensional quantum supergravity, there still exists neither clear quantization technique nor classical expansion for the membrane \cite{Horava:1995qa}. See, \emph{e.g.}, \cite{Taylor:2001vb, Dasgupta:2002iy} for excellent reviews.

According to the Matrix theory description, a supermembrane is unstable and tends to be dissolved into dynamical bits \cite{deWit:1988xki}.\,\footnote{This instability of the membrane was resolved in the Matrix theory description by realizing that the Hilbert space of the matrix quantum mechanics naturally contains multi-particle states of the D0-bits \cite{Banks:1996vh}.} This suggests that membranes might not be the fundamental objects. However, it is still highly valuable to probe M-theory from the worldvolume perspectives, especially in regard of the lacking of any covariant formalism of supermembranes. Since the three-dimensional sigma model is nonrenormalizable, which suggests that new physics may arise above a particular cutoff. It is therefore natural to ask whether there exists a renormalizable QFT at our disposal, which can be regarded as an ultra-violet (UV) completion of the nonrenormalizable membrane theory. If so, there might exist a stable notion of membranes that only become strongly coupled at low energies, where the Matrix theory description becomes valid and the physics is captured by emergent multiparticle states.
Unfortunately, no such renormalizable worldvolume theory seems to be available within the relativistic framework.

It is pioneered in \cite{Horava:2008ih} that the relativistic membrane might admit a UV completion that exhibits nonrelativistic behaviors on the worldvolume: a renormalizable sigma model describing membranes propagating in spacetime can be constructed if an anisotropy between the worldvolume space and time is introduced. Furthermore, this sigma model is required to satisfy a Lifshitz scaling with the dynamical critical exponent $z=2$\,, which measures the anisotropy between the worldvolume time $t$ and space $\mathbf{x}$\,,
\be
	t \rightarrow b \, t\,,
		\qquad%
	\mathbf{x} \rightarrow b^{1/z} \, \mathbf{x}\,. 
\ee
Consequently, the worldvolume degrees of freedom satisfy a quadratic dispersion relation,
\be \label{eq:z2dr}
	\omega^2 \sim |\mathbf{k}|^4\,, 
\ee
where $\omega$ denotes the frequency and $\mathbf{k}$ the two-dimensional spatial momentum. The membrane described by such a Lifshitz-type sigma model is referred to as the \emph{membrane at quantum criticality} \cite{Horava:2008ih}. On a curved worldvolume, this Lifshitz sigma model is coupled to three-dimensional Ho\v{r}ava gravity, which geometrizes the foliation-preserving diffeomorphisms compatible with the worldvolume anisotropy. Since the worldvolume is foliated by leaves of constant time, it is possible to consistently restrict the sum over general three-manifolds in the membrane theory to foliated manifolds, where the spatial leaves are Riemann surfaces. This may facilitate a perturbative expansion in membranes, akin to the perturbative expansion with respect to the genera of Riemann surfaces in string theory \cite{Horava:2008ih}. At low energies, a relevant deformation is turned on and modifies the dispersion relation \eqref{eq:z2dr} to be 
\be
	\omega^2 \sim |\mathbf{k}|^4 + c^2 \, |\mathbf{k}|^2\,,
\ee
with a coupling $c$ that is dimensionful from the UV perspective around the $z=2$ Lifshitz point. Classically, this relevant deformation ostensibly generates a renormalization group (RG) flow toward the relativistic, nonperturbative membrane theory in the deep infrared (IR), where low-energy Lorentz symmetry emerges. Whether and how this RG flow toward a Lorentzian fixed point could be realized quantum mechanically poses a challenging question, which requires detailed analysis of the worldvolume dynamics. If achieved, it is then tempting to conjecture that the ultimately fundamental objects in M-theory might be such nonrelativistic membranes at quantum criticality.

\vspace{2mm}
 
In this paper, we study a renormalizable three-dimensional Lifshitz-type sigma model that exhibits a $z=2$ Lifshitz scaling symmetry at the critical point. This sigma model constitutes an essential ingredient for the construction of the membrane at quantum criticality. We will focus on the case where the worldvolume is flat. The full-fledged membrane theory, however, requires us to couple the matter contents of the sigma model to dynamical worldvolume Ho\v{r}ava gravity. On the other hand, in the absence of worldvolume gravitational dynamics,  the quantization of Lifshitz-type sigma models already imposes intriguing challenges and requires new techniques to tackle with. 

In the bosonic case, the most general three-dimensional sigma model at a $z=2$ Lifshitz point is coupled to higher-rank tensorial background fields, which makes the quantization rather difficult. In the simpler case where the target space is $O(N)$ symmetric, the $z=2$ nonlinear sigma model (NLSM) already exhibits intriguing RG properties \cite{Anagnostopoulos:2010gw, lnlsm, Yan:2017mse}, for which we will give a quick review as a preparation for the quantization of Lifshitz sigma models in arbitrary geometric background fields. The study of Lifshitz $O(N)$  NLSM bears potential applications to the condensed matter systems of quantum spin liquids \cite{Rokhsar:1988zz, ardonne2004topological, fradkin2013field} and quantum spherical models \cite{selke1988annni, vojta1996quantum, Gomes:2013jba}. 

In connection to the studies of supermembranes, we are ultimately interested in supersymmetric sigma models, which bring significant simplifications in comparison to the pure bosonic case. The three-dimension $\CN=1$ supersymmetric sigma model around a $z=2$ Lifshitz point has been introduced in \cite{as, ssl1} (also see \cite{Yan:2021xnp} for a review), which is power-counting renormalizable. The relevant target-space geometry is Lorentzian and has two independent metrics, but there is no extra higher-form tensorial structure as in the bosonic case. In the deep IR, a relevant deformation controlled by the coupling constant associated with the worldvolume speed of light dominates, which drives the theory toward the relativistic fixed point satisfying a $z=1$ Lifshitz scaling. We will generalize the heat kernel method in \cite{Grosvenor:2021zvq, Yan:2021xnp} to supersymmetric field theories on a foliated spacetime manifold, and use it to study the RG flows of the supersymmetric Lifshitz sigma model.  
This will form an essential step toward determining whether such a worldvolume theory is qualified to be a quantum membrane theory that UV completes the relativistic supermembranes, which also requires coupling the Lifshitz sigma model to dynamical worldvolume supergravity. At the $z=2$ Lifshitz point, the worldvolume supergravity is also power-counting renormalizable in three dimensions and possesses a preferred time direction, which makes it possible to study its quantization as a conventional quantum field theory.

\vspace{2mm}

The paper is organized as follows. In Section~\ref{sec:nsmons}, as a warm-up, we discuss the renormalization of three-dimensional $z=2$ Lifshitz $O(N)$ NLSM. In Section~\ref{sec:gsmlp}, we study three-dimensional $\CN=1$ supersymmetric $z=2$ Lifshitz sigma model, which is described by the action \eqref{eq:ssusyz2} and is coupled to a bimetric target-space geometry. We derive the one-loop beta-functionals for the bimetric fields in this Lifshitz sigma model in Eq.~\eqref{eq:betaGH}, which form the major results of this paper. In Section~\ref{sec:qcm}, as a preliminary step toward the formulation of the supersymmetric quantum critical membrane, we couple the bosonic part of the sigma model to dynamical worldvolume Ho\v{r}ava gravity. We conclude the paper in Section~\ref{sec:concl}.

\section{\texorpdfstring{$O(N)$}{ON} Nonlinear Sigma Models at a \texorpdfstring{$z=2$}{z2} Lifshitz point} \label{sec:nsmons}

In this section, we illustrate some of the basic ingredients in our later construction of general three-dimensional Lifshitz sigma models by a simple example, where the target space of the sigma model is $O(N)$ symmetric. We parametrize the target space by a vector field $n^a$\,, $a = 1, \cdots, N$\,, which satisfies the constraint
\be \label{eq:nconst}
	n \cdot n = 1\,.
\ee
We will first review the relativistic $O(N)$ NLSM in three dimensions, which is \emph{not} renormalizable. Then, we proceed to the construction of the three-dimensional $O(N)$ NLSM at a $z=2$ Lifshitz point, which flows toward the relativistic NLSM in the deep IR.

\subsection{Relativistic  \texorpdfstring{$O(N)$}{ON} Nonlinear Sigma Model}

In the case where the three-dimensional worldvolume is relativistic, the $O(N)$ NLSM can be described by the following action:
\be \label{eq:relon}
	S^{}_\text{rel} = \frac{1}{2 \, g_\text{rel}^2} \int d^3 x \, \p^\alpha n \cdot \p_\alpha n\,.
\ee
Here, we have introduced the worldvolume coordinates $x^\alpha = (x^0 = t\,, x^i)$\,, $i = 1, \, 2$\,. The action \eqref{eq:relon} needs to be supplemented by the constraint \eqref{eq:nconst}. Instead of using the redundant parametrization in terms of $n^a$ and imposing the constraint \eqref{eq:nconst} by hand, from the QFT perspective, it is more convenient to instead adopt an intrinsic parametrization, with
\be \label{eq:ipnx}
	n = (X^1, \cdots, X^{N-1}, \, \sigma)\,,
		\qquad%
	\sigma = \sqrt{1 - X \cdot X}\,,
\ee 
where $X^I$, $I = 1, \cdots, N-1$ represent a local coordinate system. In terms of the intrinsic coordinates $X^I$, the original action \eqref{eq:relon} can be rewritten as
\be \label{eq:srelx}
	S^{}_\text{rel} = \frac{1}{2 \, g_\text{\text{rel}}^2} \int d^3 x \, \p^\alpha X^I \, \p_\alpha X^J \, G^{}_{IJ} (X)\,,
\ee
where we introduced the round metric,
\be \label{eq:rm}
	G^{}_{IJ} = \delta^{}_{IJ} + \frac{X^I \, X^J}{1- X \! \cdot \! X}\,.
\ee
The independent fields $X^I$ are Nambu-Goldstone bosons from spontaneously breaking the symmetry group from $O(N)$ to $O(N-1)$\,. In other words, the $O(N)$ global symmetry is nonlinearly realized in terms of $X^I$\,. Performing the rescaling $X^I \rightarrow g^{}_\text{rel} \, X^I$ gives
\be \label{eq:srelxexp}
	S_\text{rel} = \frac{1}{2} \int d^3 x \, \Bigl[ \p^\alpha X \cdot \p_\alpha X + \sum_{n=0}^\infty g_\text{rel}^{2(n+1)} \, \bigl( X \cdot X \bigr)^n \bigl( X \cdot \p^\alpha X \bigr) \bigl( X \cdot \p_\alpha X \bigr) \Bigr]\,.
\ee
The engineering scaling dimensions are
\be
	[x^\alpha] = -1\,,
		\qquad%
	[X^I] = \frac{1}{2}\,,
		\qquad%
	[g^{}_\text{rel}] = - \frac{1}{2}\,.
\ee
The coupling constant $g^{}_\text{rel}$ has a negative dimension, implying that the action \eqref{eq:srelxexp} is \emph{not} power-counting renormalizable. If the Wilsonian picture holds, then new physics has to appear at the energy scale $\sim 1 / g_\text{rel}^2$\,. 

\subsection{Scalar Field Theories at a Lifshitz Point}

At high energies, it is possible to formulate a renormalizable NLSM exhibiting a Lifshitz scaling symmetry, such that it flows toward the relativistic $O(N)$ NLSM described by the action \eqref{eq:srelx} in the deep IR. We build our UV theory around a Gaussian fixed point satisfying the following symmetry principles:
\begin{itemize}

\item
	
	\emph{Aristotelian spacetime symmetries}, consisting of a time translation, spatial translations, and a spatial rotation, parametrized by $\zeta$\,, $\xi^i$\,, and $\omega^{}_{ij} = \omega \, \epsilon^{}_{ij}$\,, respectively, with
\begin{align} \label{eq:adiffeo}
	\delta t = \zeta\,,
		\qquad%
	\delta {x}^i = \xi^i + \omega^{i}{}^{}_j \, {x}^j\,.
\end{align}   
Spacetime satisfying the above isometries, but without any boosts (neither Lorentzian nor Galilean), is referred to as \emph{Aristotelian spacetime} \cite{Penrose:1968ar}.\,\footnote{The notion of ``Aristotelian spacetime" can be dated back to A.~Trautman's work. Also see \cite{Grosvenor:2016gmj} where this terminology is revived.} Such spacetime is equipped with a codimension-one foliation structure and an absolute notion of time, with the spatial slices being the leaves in the foliation.

\item 
	
	\emph{Lifshitz scaling symmetry}, under which the temporal and spatial coordinates scale anisotropically, with
\be
	t \rightarrow b \, t\,,
		\qquad%
	\mathbf{x} \rightarrow b^{1/z} \, \mathbf{x}\,,
\ee
where $z$ is referred to as the dynamical critical exponent. We thus assign the following scaling dimensions to $t$ and $\mathbf{x}$\,:
\be
	[t] = -1\,,
		\qquad%
	[\mathbf{x}] = - 1/z\,,
\ee
such that energy has dimension one.

\end{itemize}
In $(D+1)$-dimensions, we consider a single real scalar field theory at the UV Gaussian fixed point with an integer $z$\,, which is described by the following free action:
\be \label{eq:fas0}
	S_0 = {\frac{1}{2}} \int dt \, d^D \mathbf{x} \, \Bigl( \p^{}_t \phi \, \p^{}_t \phi - \zeta_z^2 \, \p^{}_{i_1} \cdots \p^{}_{i_z}  \phi \, \p^{}_{i_1} \cdots \p^{}_{i_z} \phi \Bigr)\,,
\ee
where we imposed the na\"{i}ve time-reversal symmetry $t \rightarrow - t$ and an additional parity symmetry such that the action is invariant under $\phi \rightarrow -\phi$\,. 
At the RG fixed point, the scaling dimension of $\phi$ is
\be
	[\phi] = \frac{D-z}{2 \, z}\,,
\ee
which $\zeta^{}_z$ is dimensionless. 
When $D=z$\,, the field $\phi$ is dimensionless and the theory reaches its lower critical dimension. For example, string theory in flat spacetime is built upon the action \eqref{eq:fas0} at the lower critical dimension with $D = z = 1$\,. This is essential for the preferred role of the strings, in contrast to general relativistic $p$-branes. Similarly, for the membrane where $D=2$\,, the theory with $z=2$ plays a special role and deserves particular attentions \cite{Horava:2008ih}. Aiming for applications to the membrane theory, we now focus on the Gaussian fixed point in (2+1)-dimensions described by the free action
\be \label{eq:fas0l}
	S_0 = \frac{1}{2} \int dt \, d^2 \mathbf{x} \, \Bigl( \p^{}_t \phi \, \p^{}_t \phi - \zeta_2^2 \, \p^{}_{i} \p^{}_{j}  \phi \, \p^{}_{i} \p^{}_{j} \phi \Bigr)\,.
\ee
When generic interactions are turned on, relevant quadratic derivative terms will be generated by quantum corrections, which deform the action \eqref{eq:fas0l} to be
\be \label{eq:iacm}
	S = \frac{1}{2} \int dt \, d^2 \mathbf{x} \, \Bigl( \p^{}_t \phi \, \p^{}_t \phi - \zeta_2^2 \, \p^{}_{i} \p^{}_{j}  \phi \, \p^{}_{i} \p^{}_{j} \phi - c^2 \, \p^{}_i \phi \, \p^{}_i \phi - m^2 \phi^2 + \cdots \Bigr)\,.
\ee   
Here, $\zeta^{}_2$ is dimensionless, while the ``speed of light" $c$ and mass $m$ have the scaling dimensions
\be
	[c] = 1/2\,,
		\qquad%
	[m] = 1\,.
\ee 
It is interesting to note that the theory \eqref{eq:iacm} is already IR finite when a nonzero $c$ is included. 
At the UV fixed point defined by Eq.~\eqref{eq:fas0l}, where $c$ and $m$ are tuned to zero, a quadratic shift symmetry emerges, with \cite{Griffin:2013dfa}
\be
	\phi \rightarrow \phi + a^{}_{ij} \, x^i \, x^j + a^{}_i \, x^i + a\,.
\ee
This symmetry protects the $z=2$ Lifshitz point from being deformed by the relevant terms at the quantum level,
and it has been generalized to higher-degree polynomial shift symmetries \cite{Griffin:2013dfa, Hinterbichler:2014cwa, Griffin:2014bta, Griffin:2015hxa} that protect a hierarchy of QFTs at Lifshitz points with different $z$'s, with the associated Gaussian theory defined by Eq.~\eqref{eq:fas0}. 

In the deep IR, the two-derivative quadratic term $\p_i \phi \, \p_i \phi$ will dominate over the four-derivative quadratic term in Eq.~\eqref{eq:iacm} and drive the theory toward a relativistic fixed point. The low-energy observers would find it natural to take a set of rescaled coordinates due to their relativistic prejudice \cite{Grosvenor:2016gmj},
\be \label{eq:rescalingc}
	y^0 = t\,,
		\qquad
	y^i = x^i / c\,,
		\qquad
	\Phi = c \, \phi\,,
\ee
with the scaling dimensions $[y^0] = [y^i] = - 1$ and $[\Phi] = 1/2$\,.
According to the low-energy relativistic observers, the action \eqref{eq:iacm} becomes 
\be 
	S = - \frac{1}{2} \int d^3 y \, \Bigl( \p_\mu \Phi \, \p^\mu \Phi + m^2 \Phi^2 + \tilde{\zeta}_2^{\,2} \, \p^{}_{i} \p^{}_{j}  \Phi \, \p^{}_{i} \p^{}_{j} \Phi + \cdots \Bigr)\,,
\ee   
where $\tilde{\zeta}^{}_2 = \zeta^{}_2 / c^2$
has the scaling dimension $[\tilde{\zeta}^{}_2] = - 1$\,. This implies that $\tilde{\zeta}^{}_2 \sim \Lambda^{-1}$\,, where $\Lambda$ is the energy cutoff. Therefore, at low energies, the four-derivative term with the coupling $\tilde{\zeta}^{}_2$ is suppressed, and an accidental Lorentz symmetry is developed. 

\subsection{\texorpdfstring{$O(N)$}{ON} Nonlinear Sigma Model at a  \texorpdfstring{$z=2$}{z2} Lifshitz Point} \label{sec:onnlsmz2}

We are now ready to apply the ingredients reviewed in the previous subsections to build an $O(N)$ NLSM at a $z=2$ Lifshitz point \cite{Anagnostopoulos:2010gw, lnlsm, Yan:2017mse}. The kinetic part of the action is
\be \label{eq:skin}
	S^{}_\text{kin.} = \frac{1}{2 \, g^2} \int dt \, d^2 \mathbf{x} \, \p^{}_t n \cdot \p^{}_t n\,,
\ee
where the vector field $n^a$ satisfies the constraint \eqref{eq:nconst}. At the RG fixed point where a $z=2$ Lifshitz scaling is developed, $n^a$ is dimensionless. The potential terms contain all possible marginal operators with four spatial derivatives, 
\be \label{eq:spot}
	S^{}_\text{pot.} = - \frac{1}{2 \, g^2} \int dt \, d^2 \mathbf{x} \, \zeta^2 \, \biggl[ \p^{}_i \p^{}_j n \cdot \p^{}_i \p^{}_j n + \frac{\eta^{\phantom{\dagger}}_1}{4} \, \bigl( \p^{}_i n \cdot \p^{}_i n \bigr)^2 + \frac{\eta^{\phantom{\dagger}}_2}{2} \, \bigl( \p^{}_i n \cdot \p^{}_j n \bigr) \bigl( \p^{}_i n \cdot \p^{}_j n \bigr) \biggr]\,,
\ee
where we introduced the marginal couplings $g$\,, $\zeta$\,, $\eta^{\phantom{\dagger}}_1$\,, and $\eta^{\phantom{\dagger}}_2$\,.
Finally, there is also a relevant deformation,
\be \label{eq:sdef}
	S^{}_\text{r.d.} = - \frac{c^2}{2 \, g^2} \int dt \, d^2 \mathbf{x} \, \p^{}_i n \cdot \p^{}_i n \,.
\ee 
The total action is therefore the sum of Eqs.~\eqref{eq:skin}~$\sim$~\eqref{eq:sdef}. As desired, in the deep IR, the relevant term \eqref{eq:sdef} will dominate over the marginal potential terms in Eq.~\eqref{eq:spot}, driving the theory toward the $z=1$ fixed point, where Lorentz symmetry emerges. In this sense, our Lifshitz sigma model acts as a UV completion of the relativistic theory \eqref{eq:relon}. Reparametrizing this Lifshitz sigma model in terms of the independent target-space coordinates $X^I$ as in Eq.~\eqref{eq:ipnx}, we find the following action:
\begin{align} \label{eq:lona}
\begin{split}
	S & = \frac{1}{2 \, g^2} \int dt \, d^2 \mathbf{x} \, \biggl\{ \Bigl( \p^{}_t X^I \, \p^{}_t X^J - \zeta^2 \, \Box X^I \, \Box X^J - c^2 \, \p^{}_i X^I \, \p^{}_i X^J \Bigr) \, G_{IJ} \\[4pt] 
	& \quad\,\, - \frac{\zeta^2}{4} \, \Bigl[ (\eta^{\phantom{\dagger}}_1+4) \, \bigl( \p^{}_i X^I \, \p^{}_i X^J \, G^{}_{IJ} \bigr)^2
	+ 2 \, \eta^{\phantom{\dagger}}_2 \, \bigl( \p^{}_i X^I \, \p^{}_j X^J \, G_{IJ} \bigr) \bigl( \p^{}_i X^K \, \p^{}_j X^L \, G_{KL} \bigr) \Bigr] \bigg\},
\end{split}
\end{align}
where $G_{IJ}$ is the round metric given in Eq.~\eqref{eq:rm} and
\be
	\Box X^I \equiv \p^2 X^I + \Gamma^I{}^{}_{JK} \, \p^{}_i X^J \, \p^{}_i X^K\,.
\ee
Here, $\Gamma^I{}_{JK}$ is the Christoffel symbol associated with the Levi-Civita connection in the target space. Namely,
\be
	\Gamma^I{}_{JK} = \frac{1}{2} \, G^{IL} \, \Bigl( \p_J G_{KL} + \p_K G_{JL} - \p_L G_{JK} \Bigr) = X^I \, G_{JK} \,.
\ee
A sufficient condition for the theory \eqref{eq:lona} to be bounded from below is that the Lagrangian density is a sum of complete squares, which requires \cite{lnlsm, Yan:2017mse}
\be \label{eq:unitaritybound}
	\eta^{}_2 \geq - 4\,, 
		\qquad%
	\eta^{}_1 + \eta^{}_2 \geq -4\,,
		\qquad%
	\eta^{}_1 + 2 \, \eta^{}_2 \geq - 4\,.
\ee

The theory \eqref{eq:lona} is power-counting renormalizable. Its associated RG flows have been studied in \cite{Anagnostopoulos:2010gw, lnlsm}.\,\footnote{The beta-function of the coupling $g$ in Eq.~\eqref{eq:lona} was first evaluated in \cite{Anagnostopoulos:2010gw}. The complete beta-functions of the other couplings, including $\eta^{\phantom{\dagger}}_{1,2}$\,, were later computed in \cite{lnlsm} (see also \cite{Yan:2017mse}).} The one-loop beta-functions are \cite{lnlsm, Yan:2017mse}
\begin{subequations}
\begin{align}
	\beta_{g} & = -  \frac{g^3}{8\pi} \, (N-2) + O (g^5)\,,
		&%
	\beta_{\zeta^2} & = 0 + O(g^4)\,, \label{eq:betag2} \\[4pt]
	\beta_{\eta^{\phantom{2}}_1} & = \frac{g^2}{32 \pi} \, F^{}_1 (\eta_1\,, \eta_2) + O (g^4)\,,  
		&%
	\beta_{\eta^{\phantom{2}}_2} & = \frac{g^2}{32 \pi} \, F^{}_2 (\eta_1\,, \eta_2) + O (g^4)\,,
\end{align}
\end{subequations}
where
\begin{subequations}
\begin{align}
	F_1 & = (2 \, N + 3) \, \eta_1^2 + 4 \, (N+3) \, \eta^{\phantom{2}}_1 \, \eta^{\phantom{2}}_2 + (N+10) \, \eta_2^2
		+ 8 \, (N+1) \, \eta^{\phantom{2}}_1 + 48 \, \eta^{\phantom{2}}_2 - 16\,, \\[2pt]
	F_2 & = \eta_1^2 + 8 \, \eta^{\phantom{2}}_1 \, \eta^{\phantom{\dagger}}_2 + (N+10) \, \eta_2^2 + 24 \, \eta^{\phantom{2}}_1 + 8 (N+4) \, \eta^{\phantom{\dagger}}_2 + 80\,.
\end{align}
\end{subequations}
The anomalous dimensions of $X^\mu$ and $c^2$ are
\be
	\gamma^{\phantom{\dagger}}_{c^2} = \frac{g^2}{16 \pi} \Bigl[ N \, \eta^{\phantom{2}}_1 + (N+2) \, \eta^{\phantom{2}}_2 + 8 \Bigr] + O(g^4)\,,
		\qquad%
	\gamma^{\phantom{\dagger}}_X = \frac{g^2}{8\pi} \, (N-1) + O(g^4)\,. 
\ee
The detailed RG structure has been analyzed in \cite{lnlsm, Yan:2017mse},\,\footnote{Based on the beta-function of $g$\,, it is claimed in \cite{Anagnostopoulos:2010gw} that the theory \eqref{eq:lona} is asymptotically free. However, this claim is not valid as the beta-functions of $\eta^{\phantom{\dagger}}_{1,2}$ are missing in \cite{Anagnostopoulos:2010gw}. In \cite{lnlsm, Yan:2017mse}, the complete analysis of the one-loop beta-functions revealed a rather rich RG structure of this Lifshitz $O(N)$ NLSM, and it is shown that the theory is only asymptotically free along specific RG trajectories.} where it is also shown explicitly that the beta-functions are one-loop exact in the large $N$ limit. In the projected $\eta^{\phantom{\dagger}}_1$-$\eta^{\phantom{\dagger}}_2$ plane, there are up to four fixed points, only one of which is independent of $N$\,. This universal fixed point is given by
\be \label{eq:dbc}
	\eta^{\phantom{\dagger}}_1 = - 4\,, 
		\qquad%
	\eta^{\phantom{\dagger}}_2 = 0\,.
\ee
This condition saturates the unitarity bound \eqref{eq:unitaritybound}, and also selects a RG trajectory that is asymptotically free if $N>2$\,. Plugging \eqref{eq:dbc} into the action \eqref{eq:lona}, we find,
\be \label{eq:dbca}
	S = \frac{1}{2 \, g^2} \int dt \, d^2 \mathbf{x} \, \Bigl( \p^{}_t X^I \, \p^{}_t X^J - \zeta^2 \, \Box X^I \, \Box X^J - c^2 \, \p^{}_i X^I \, \p^{}_i X^J \Bigr) \, G^{}_{IJ}\,.
\ee
When $c$ is tuned to zero, the potential part of the action \eqref{eq:dbca} is a complete square of $\Box X^I$ and the theory is at the ``detailed balance" \cite{Horava:2008ih},  which is closely related to the effective action of stochastic quantization \cite{Parisi:1980ys, namiki2008stochastic} with the Parisi-Sourlas supersymmetry \cite{Parisi:1979ka, SOURLAS1985115}.\,\footnote{Also see \cite{damgaard1987stochastic} for a review and,  \emph{e.g.}, \cite{zinn1986renormalization} for an application to sigma models.} At equilibrium, the theory is effectively described by a two-dimensional Euclidean NLSM. Generically, the ground state wave functions of a theory at the detailed balance are given by the partition function of a Euclidean field theory in one dimension lower. In the next subsection, we will see that Eq.~\eqref{eq:dbca} is the bosonic part of a Lifshitz NLSM with $\CN=1$ supersymmetry. 

Along other lines, it is also interesting to note that this type of Lifshitz $O(N)$ NLSMs naturally arise from condensed matter physics, such as the (generalized) Rokhsar-Kivelson (RK) models, in which context the detailed balance condition \eqref{eq:dbc} is known as the RK point. In the original RK quantum dimer model that describes the quantum spin liquid, the ground state is given by the equal superposition of all dimer configurations \cite{Rokhsar:1988zz, ardonne2004topological, fradkin2013field}. Furthermore, the precise notion of the $O(N)$ Lifshitz NLSM \eqref{eq:lona} appears to be closely related to quantum spherical models with competing interactions, which are used to describe critical phenomena and phase transitions in condensed matter systems such as the axial next-nearest neighbor Ising (ANNNI) model \cite{selke1988annni, vojta1996quantum, Gomes:2013jba}. The ANNNI model exhibits intriguing ordered, disordered, and modulated phases that meet at a Lifshitz point. 

\subsection{Lifshitz Nonlinear Sigma Model with \texorpdfstring{$\CN=1$}{N1} Supersymmetry} \label{sec:onn1susy}

In the previous subsection, we have seen that the $O(N)$ NLSM at a $z=2$ fixed point is already rather complicated, due to the extra four-derivative interactions with the couplings $\eta^{\phantom{\dagger}}_1$ and $\eta^{\phantom{\dagger}}_2$\,. Intriguingly, there exists an RG fixed point \eqref{eq:dbc} on the projected $\eta^{\phantom{\dagger}}_1$-$\eta^{\phantom{\dagger}}_2$ plane that leads to a fairly simple theory \eqref{eq:dbca}, which is stable and asymptotically free. In this subsection, following \cite{ssl1, as, Gomes:2016tus}, we show that this theory is the bosonic part of a supersymmetric NLSM. 

In three dimensions, the $\CN=1$ real supercharges are $Q_\alpha = ( Q_-\,, Q_+)$\,, which are Grassmannian. To define the supersymmetry algebra in three-dimensional Aristotelian spacetime, we first introduce the Dirac gamma matrices,
\be
	\rho^0 = 
	\begin{pmatrix}
		0 &\,\, 1 \\
		-1 &\,\, 0
	\end{pmatrix}\,,
		\qquad%
	\rho^1 = 
	\begin{pmatrix}
		0 &\,\,\,\, 1 \\
		1 &\,\,\,\, 0
	\end{pmatrix}\,,
		\qquad%
	\rho^2 = 
	\begin{pmatrix}
		1 &\,\, 0 \\
		0 &\,\, -1
	\end{pmatrix}\,.
\ee 
For any given Grassmannian variable $\chi^{}_\alpha$\,, its conjugate is defined to be $\bar{\chi} = i \, \chi^\intercal \, \rho^0$. We also define $\chi^2 \equiv \bar{\chi} \, \chi$\,. The isometry group for Aristotelian spacetime is generated by the time translation $P^{}_0$\,, spatial translations $P^{}_i$\,, and spatial rotations $J^{}_{ij} = \tfrac{1}{2} \, J \, \epsilon^{}_{ij}$\,. These generators are associated with the Lie algebra parameters $\zeta$\,, $\xi^i$\,, and $\omega^i{}_j$ in \eqref{eq:adiffeo}, respectively. We define super-Aristotelian algebra via the following non-vanishing Lie brackets:
\begin{align} \label{eq:saa}
	[J\,, Q] = i \, \rho^0 \, Q\,,
		\qquad%
	[P^{}_i\,, J] = 2 \, \epsilon^{}_{i}{}^j \, P^{}_j\,,
		\qquad%
	\{ Q\,, \bar{Q} \} = 2 \, i \, \slashed{P}\,.
\end{align}
where $\slashed{P} = \rho^0 \, P_0 + c \, \rho^i \, P_i$\,.
This super-Aristotelian algebra can be extended by introducing the Lorentz boost generators $J_{0i}$ to the well-known super-Poincar\'{e} algebra\,. 

In the superspace formalism,\,\footnote{See also the superspace formalism in \cite{Auzzi:2019kdd, Baiguera:2022cbp} for Galilean-invariant supersymmetric models.} we introduce a pair of Grassmannian coordinates $\theta_\alpha$\,. Measured in energy, the scaling dimensions are
\be
	[t] = -1\,,
		\qquad
	[x^i] = [\theta_\alpha] = - \frac{1}{2}\,.
\ee
The symmetry algebra \eqref{eq:saa} can be realized by introducing the following operators associated with the generators $P_0$\,, $P_i$\,, $J$\,, and $Q$\,, respectively:
\be
	\CP^{}_0 = - i \, \p^{}_0\,,
		\qquad%
	\CP^{}_i = - i \, \p^{}_i\,,
		\qquad%
	\CJ = 2 \, i \, \epsilon^{ij} \, x^{}_i \, \p^{}_j - \bar{\theta} \, \frac{\p}{\p \theta}\,,	
		\qquad%
	\CQ = \frac{\p}{\p \bar{\theta}} - \slashed{\p} \, \theta \,,
\ee
where $\slashed{\p} = \rho^0 \, \p_t + c \, \rho^i \, \p_i$\,.
Furthermore, we define a superfield $Y^I = Y^I (t\,, x^i\,, \theta_\alpha)$\,. %
Expanding the superfield with respect to $\theta_\alpha$\,, we find
\be
	Y^I (t\,, x^i, \theta_\alpha) = X^I ( t\,, x^i) + \bar{\theta} \, \Psi^I (t\,, x^i) + \tfrac{1}{2} \, \theta^2 \, B^I (t\,, x^i)\,,
\ee
where $\Psi^I_\alpha$ is a vector of two-spinors and $B^I$ is an auxiliary field.
The supersymmetry transformation of the superfield $Y^I$ parametrized by $\epsilon = (\epsilon_-\,, \epsilon_+)^\intercal$ is given by
\be
	\delta_\epsilon Y^I = [\bar{\epsilon} \, \CQ\,, Y^I]\,, 
\ee
In components, we have
\begin{align}
	\delta_\epsilon X^I = \bar{\epsilon} \, \Psi^I\,,
		\qquad%
	\delta_\epsilon \Psi^I = \bigl( - \slashed{\p} X^I + B^I \bigr) \, \epsilon\,, 
		\qquad%
	\delta_\epsilon B^I = \bar{\epsilon} \, \slashed{\p} \Psi^I\,,
\end{align}
The supercovariant derivative that commutes with $\CQ$ is
\be \label{eq:covder}
	D = \frac{\p}{\p \bar{\theta}} + \slashed{\p} \, \theta\,,
		\qquad%
	\text{s.t.} \,\, \{ \CQ, D \} = 0\,.
\ee
The supersymmetric $O(N)$ sigma model at a $z=2$ Lifshitz point is 
\begin{align} \label{eq:susyon}
	S_\text{susy} = \frac{1}{2 \, g^2} \int dt \, d^2 \mathbf{x} \, d^2 \theta \, \Bigl( \bar{D}^\alpha Y^I \, D_\alpha Y^J - 2 \, \zeta \, \p^{}_i Y^I \, \p^{}_i Y^J \Bigr) \, G_{IJ} (Y)\,,
\end{align}
where $G_{IJ} (Y)$ is the round metric that generalizes Eq.~\eqref{eq:rm}, with
\be
	G_{IJ} (Y) = \delta_{IJ} + \frac{Y^I \, Y^J}{1- Y \cdot Y}\,.
\ee
The action \eqref{eq:susyon} supersymmetries the bosonic action \eqref{eq:dbca} at the detail balance. The one-loop beta-functions of the two marginal couplings $g$ and $\zeta$ are given by
\be \label{eq:betagsusy}
	\beta_{g} = -  \frac{g^3}{8\pi} \, (N-2) + O (g^5)\,,
		\qquad%
	\beta_{\zeta^2} = 0 + O(g^4)\,,
\ee
which coincide with Eq.~\eqref{eq:betag2} for the bosonic case. 

From the perspective of a low-energy relativistic observer, it is convenient to rewrite the action \eqref{eq:susyon} using the following rescalings analogous to Eq.~\eqref{eq:rescalingc}:
\be 
	y^0 = t\,,
		\qquad
	y^i = x^i / c\,,
		\qquad
	\CY^i = c \, g^{-1} \, Y^i\,.
\ee
As a result, the action \eqref{eq:susyon} becomes
\begin{align} \label{eq:rssusya}
	S_\text{susy} = \frac{1}{2} \int d^3 y \, d^2 \theta \, \Bigl( \bar{D}^\alpha \CY^I \, D_\alpha \CY^J - 2 \, \tilde{\zeta} \, \p^{}_i \CY^I \, \p^{}_i \CY^J \Bigr) \, G_{IJ} (g_\text{r} \, \CY)\,,
\end{align}
where 
\be
	g_\text{r} = g / c\,, 
		\qquad%
	\tilde{\zeta} = \zeta / c^2\,,
		\qquad%
	D = \frac{\p}{\p \bar{\theta}} + \rho^0 \, \theta \, \frac{\p}{\p y^0} + \rho^i \, \theta \, \frac{\p}{\p y^i}\,.
\ee
We also defined
\be
	\int d^2 \theta \, \delta^{(2)} (\theta) = 1\,,
		\qquad%
	\delta^{(2)} (\theta) = \theta^2\,. 
\ee
As $[ \, \tilde{\zeta} \, ] = -1$\,, the associated term is suppressed compared to the first term in Eq.~\eqref{eq:rssusya}, leading to the relativistic supersymmetric $O(N)$ nonlinear sigma model,
\be \label{eq:susysrel}
	S_\text{rel} = \frac{1}{2} \int d^3 y \, d^2 \theta \, \bar{D}^\alpha \CY^I \, D_\alpha \CY^J \, G_{IJ} (g^{}_\text{r} \, \CY)\,.
\ee
Here, $[g^2_\text{r}] = -1$ and the theory is non-renormalizable. The action \eqref{eq:susysrel} supersymmetrizes the bosonic action \eqref{eq:srelx}, up to a rescaling of the embedding coordinates.

\section{General Sigma Models at a \texorpdfstring{$z=2$}{z2} Lifshitz Point} \label{sec:gsmlp}

In this section, we consider general $z=2$ Lifshitz sigma models in three dimensions, coupled to arbitrary background fields. We will first discuss the classical bosonic theory before introducing the $\CN = 1$ supersymmetric Lifshitz sigma model. This supersymmetric sigma model is coupled to a bimetric target-space geometry \cite{as, ssl1}. We then develop the background field method and a covariant heat kernel approach for deriving the one-loop effective action associated with the supersymmetric Lifshitz sigma model, from which we will derive the beta-functionals of the target-space metric fields. 

\subsection{The Classical Theory} \label{sec:ct}

The nonlinear sigma model with a global $O(N)$ symmetry and a dynamical critical exponent $z=2$ has been given in Eq.~\eqref{eq:lona}. Relaxing this $O(N)$ symmetry, a sigma model in arbitrary background fields can be constructed. Consider the worldvolume field $X^I$\,, $I = 1, \cdots, d$ that maps the three-dimensional worldvolume to the $d$-dimensional target space. We assume that the signature of the target-space manifold is Euclidean. The most general bosonic sigma model around a $z=2$ Lifshitz point is 
\begin{align} \label{eq:bgsmz2}
\begin{split}
	S & = \frac{1}{2 \, g^2} \int \! dt \, d^2 \mathbf{x} \, \Bigl[ \p^{}_t X^I \, \p^{}_t X^J \, G^{}_{IJ} (X) - \Box X^I \, \Box X^J \, E^{}_{IJ} (X) - c^2 \, \p^{}_i X^I \, \p^{}_i X^J \, \tilde{G}^{}_{IJ} (X) \\[4pt] 
	& \hspace{5.5cm} - m^2 \, M (X) - \p^{}_i X^I \, \p^{}_i X^J \, \p^{}_j X^K \, \p^{}_j X^L \, T^{}_{IJKL} (X) \Bigr]\,,
\end{split}
\end{align}
Here, $G_{IJ}$\,, $\tilde{G}_{IJ}$ and $E_{IJ}$ are symmetric two-tensors and $T_{IJKL}$ is a four-tensor, satisfying the conditions $T_{IJKL} = T^{}_{JIKL} = T^{}_{IJLK} = T^{}_{KLIJ}$\,.  
We also introduced an affine connection $\Theta$ such that
\be \label{eq:boxx}
	\Box X^I = \p^2 X^I \! + \Theta^I{}^{}_{JK} \, \p^{}_i X^J \, \p^{}_i X^K\,.
\ee
Moreover, a na\"{i}ve time-reversal symmetry $t \rightarrow - t$ has been imposed, such that the Wess-Zumino term,
\be
	S_\text{WZ} \sim \int dt \, d^2 \mathbf{x} \, \epsilon^{\alpha\beta\gamma} \, \p_\alpha X^\mu \, \p_\beta X^\nu \, \p_\gamma X^\rho \, A_{\mu\nu\rho}\,,
\ee
is forbidden. Here, $A_{\mu\nu\rho}$ is a three-form gauge field, which is expected to play an important role when the sigma model is promoted to be a membrane theory. In the deep IR around the relativistic fixed point, $A_{\mu\nu\rho}$ would correspond to the three-form gauge field in eleven-dimensional supergravity. In this paper, for simplicity, we will assume the na\"{i}ve time-reversal symmetry and therefore \emph{not} include this three-form gauge field.  

In the following, we consider three-dimensional $\CN=1$ supersymmetric Lifshitz sigma model in general background fields. This type of sigma models have been introduced in \cite{as, ssl1}, whose target-space geometry turns out to have a bimetric structure. The bosonic part of the supersymmetric bimetric sigma model will exhibit significant simplification compared to the bosonic action \eqref{eq:bgsmz2}, in the way that the four-tensor structure is prohibited by the Aristotelian supersymmetry.

\subsubsection{Supersymmetric sigma models and bimetricity}

Now, following closely \cite{as, ssl1}, we consider the supersymmetric generalization of the bosonic sigma model \eqref{eq:bgsmz2}, using the ingredients from Section~\ref{sec:onn1susy}. In arbitrary background fields, the $\CN=1$ supersymmetric $O(N)$ Lifshitz NLSM \eqref{eq:susyon} is generalized to be \cite{as, ssl1}
\begin{align} \label{eq:ssusyz2}
	S_\text{susy} = \frac{1}{2 \, g^2} \int dt \, d^2 \mathbf{x} \, d^2 \theta \, \Bigl[ \bar{D}^\alpha Y^I \, D_\alpha Y^J \, G^{}_{IJ} (Y) - 2 \, \p^{}_i Y^I \, \p^{}_i Y^J \, H^{}_{IJ} (Y) - 2 \, m^2 \, U(Y) \Bigr] \,,
\end{align}
where $G_{IJ}$ and $H_{IJ}$ are symmetric two-tensors that play the role of target-space metrics. For now, we require that the target-space be a Riemannian (instead of pseudo-Riemannian) manifold, such that the signatures of both $G_{IJ}$ and $H_{IJ}$ are positive. Due to the lack of any boost symmetry on the worldvolume, \emph{a priori}, these two background fields are \emph{not} related to each other. This guarantees that the theory is unitary. Disregarding the mass term, which is relevant around the $z=2$ Lifshitz point and becomes suppressed in the UV, the target space corresponding to the worldvolume action \eqref{eq:ssusyz2} develops an intriguing \emph{bimetric} feature. This is reminiscent of the two-dimensional Lifshitz sigma model considered in \cite{Yan:2021xnp}, which is in contrast purely bosonic.  

It is illuminating to write down the full $\CN=1$ supersymmetric action after integrating out $\theta_\alpha$ in Eq.~\eqref{eq:ssusyz2}. For this purpose, we introduce $\p_\mu = (\p_t\,, c \, \p_i)$\,, together with the Minkowski metric $\eta^{\mu\nu} = \text{diag} (-1, \, 1, \, 1)$\,. Moreover, we define the Christoffel symbols associated with the Levi-Civita connections $\nabla$ for the metric field $G_{IJ}$\,, 
\begin{subequations} \label{eq:csg}
\begin{align} 
	\Gamma^I{}^{}_{JK} & = \frac{1}{2} \, G^{IL} \Bigl( \p^{}_J G^{}_{KL} + \p^{}_K G^{}_{JL} - \p^{}_L G^{}_{JK} \Bigr)\,, \label{eq:csgamma} 
\end{align}
and the Levi-Civita connection $\Delta$ for the metric field $H_{IJ}$\,,
\begin{align} 
	\Theta^I{}^{}_{JK} & = \frac{1}{2} \, H^{IL} \Bigl( \p^{}_J H^{}_{KL} + \p^{}_K H^{}_{JL} - \p^{}_L H^{}_{JK} \Bigr)\,. \label{eq:csgt}
\end{align}
\end{subequations}
The difference between the Christoffel symbols define a (1,\,2)-tensor,
\be \label{eq:sijk}
	S^I{}^{}_{JK} = \Gamma^I{}^{}_{JK} - \Theta^I{}^{}_{JK}\,.
\ee
We also define the Riemann tensors $R^I{}^{}_{JKL}$ for $G^{}_{IJ}$\,,
\begin{subequations} \label{eq:rsigma}
\begin{align}
	R^I{}^{}_{JKL} & = \p^{}_K \Gamma^I{}^{}_{JL} - \p^{}_L \Gamma^I{}^{}_{JK} + \Gamma^I{}^{}_{KM} \, \Gamma^M{}^{}_{JL} - \Gamma^I{}^{}_{LM} \, \Gamma^M{}^{}_{JK}\,, 
\end{align}
and $\Sigma^I{}_{JKL}$ for $H_{IJ}$\,,
\begin{align}
	\,\,\,\,\,\Sigma^I{}^{}_{JKL} & = \p^{}_K \Theta^I{}^{}_{JL} - \p^{}_L \Theta^I{}^{}_{JK} + \Theta^I{}^{}_{KM} \, \Theta^M{}^{}_{JL} - \Theta^I{}^{}_{LM} \, \Theta^M{}^{}_{JK}\,.
\end{align}
\end{subequations}
Performing the integral over $\theta_\alpha$ in Eq.~\eqref{eq:ssusyz2}, and in terms of the above geometric ingredients, we find that the following $\CN=1$ supersymmetric action: 
\begin{align} \label{eq:ssusycomp}
\begin{split}
	S_\text{susy} = \frac{1}{2 \, g^2} \! \int \! dt \, d^2 \mathbf{x} \, \biggl\{ \! & - \Bigl( \, \p^\mu X^I \p_\mu X^J + \overline{\Psi}{}^I \slashed{\nabla} \Psi^J \Bigr) \, G_{IJ} (X) \\[2pt]
		& + \tfrac{1}{8} \, \overline{\Psi}{}^I \, \Psi^J \, \overline{\Psi}{}^K \, \Psi^L \, G_{IM} (X) \, R^M{}_{KJL} (X) \\[4pt]
		& - \Big( \, \Box X^I \, \Box X^J \! + \overline{\Psi}{}^I \Box \Psi^J \Bigr) \, H_{IK} (X) \, G^{KL}(X) \, H_{LJ} (X) \\[4pt]
		& + \overline{\Psi}{}^I \, \Psi^J \Bigl[ \, \Box X^K \, S^M{}_{IJ} (X) + \p^i X^K \, \p_i X^L \, \Sigma^M{}_{ILJ} (X) \, \Bigr] \, H_{MK} (X) \\[2pt]
		& - m^2 \, \Bigl[ \, W (X) - \tfrac{1}{2} \, \overline{\Psi}{}^I \, \Psi^J \, \nabla^{}_{\!I} \nabla^{}_{\!\!J} U (X) \, \Bigr] \biggr\}\,.
\end{split}
\end{align}
Here, $\Box = \Delta_{i} \, \Delta_i$\,, and $\Box X^I$ has been defined in Eq.~\eqref{eq:boxx}, where $\Theta$ is the Levi-Civita connection associated with the metric field $H_{IJ}$ and is defined in Eq.~\eqref{eq:csgt}. Moreover, 
\be
	W (X) = - \p^{}_I U(X) \, G^{IJ}(X) \, H^{}_{JK}(X) \, \Box X^K + \frac{1}{4} \, m^2 \, \p^{}_I U(X) \, G^{IJ}(X) \, \p^{}_{\!J} U(X)\,.
\ee
In particular, the bosonic part of the action \eqref{eq:ssusycomp} is given by
\begin{align} \label{eq:sbsusy}
\begin{split}
	S^\text{B}_{\text{susy}} = \frac{1}{2 \, g^2} \! \int \! dt \, d^2 \mathbf{x} \, \Bigl[ \, \p^{}_t  X^I \, \p^{}_t X^J \, G^{}_{IJ} (X) & - \Box X^I \, \Box X^J \, H^{}_{IK} (X) \, G^{KL} (X) \, H^{}_{LJ} (X) \\[2pt]
		& - c^2 \, \p^{}_i  X^I \, \p^{}_i X^J \, G^{}_{IJ} (X) - m^2 \, W(X) \, \Bigr]\,,
\end{split}
\end{align}
Note that the general bosonic action \eqref{eq:bgsmz2} reduces to Eq.~\eqref{eq:sbsusy} under the identifications in Eq.~\eqref{eq:csgt} and
\be
	E^{}_{IJ} = H^{}_{IK} \, G^{KL} \, H^{}_{LJ}\,, 
		\qquad%
	\tilde{G}^{}_{IJ} = G^{}_{IJ}\,,
		\qquad%
	M = W\,,
		\qquad%
	T^{}_{IJKL} = 0\,.
\ee

Around the relativistic fixed point, the $H_{IJ}$ term in Eq.~\eqref{eq:ssusyz2} becomes irrelevant, and the action \eqref{eq:ssusyz2} flows toward the form of Eq.~\eqref{eq:susysrel}, but now $G_{IJ}$ is not anymore restricted to be the round metric \eqref{eq:rm}. In components, the low-energy relativistic $\CN=1$ supersymmetric sigma model is given by (with $m$ tuned to zero)
\begin{align} \label{eq:ssusycomprel}
\begin{split}
	S_\text{rel.} = \frac{1}{2 \, g^2} \int dt \, d^2 \mathbf{x} \, \biggl\{ & \! - \! \Bigl( \, \p^\mu X^I \p_\mu X^J + \overline{\Psi}{}^I \slashed{\nabla} \Psi^J \Bigr) \, G^{}_{IJ} (X) \\
	& \! + \tfrac{1}{8} \, \overline{\Psi}{}^I \, \Psi^J \, \overline{\Psi}{}^K \, \Psi^L \, G^{}_{IM} (X) \, R^M{}^{}_{KJL} (X) \biggr\}\,,
\end{split}
\end{align}
which collects the terms in Eq.~\eqref{eq:ssusycomp} that are independent of $H^{}_{IJ}$\,.

\subsubsection{Covariant background field method} \label{sec:cbfm}

To facilitate the later evaluation of the one-loop quantum effective action associated with the supersymmetric sigma model \eqref{eq:ssusyz2}, we now develop the relevant background field method. We will derive the Taylor expansion of the action \eqref{eq:ssusyz2} with respect to a fluctuating vector field that is covariant with respect the background bimetric geometry. We will focus on the beta-functionals of $G_{IJ}$ and $H_{IJ}$\,, which only contain marginal couplings. We therefore tune the dimensionful couplings $c$ and $m$ to zero at the classical level.\,\footnote{The IR divergence will be cured by a regulator that we introduce later in Eq.~\eqref{eq:olea}.}    
It is convenient to first perform a Wick rotation by setting $t = - i \, \tau$\,, such that 
\be \label{eq:se}
	S^{}_\text{E} = \frac{1}{2} \int d\tau \, d^2 \mathbf{x} \, d^2 \theta \, \Bigl( \bar{D}^\alpha Y^I \, D_\alpha Y^J \, G^{}_{IJ} + 2 \, \p^{}_i Y^I \, \p^{}_i Y^J \, H^{}_{IJ} \Bigr)\,,
\ee
where 
\be
	D_\alpha = - i \, \frac{\p}{\p \bar{\theta}^\alpha} + \bigl( \rho^0 \, \theta \bigr)_\alpha \, \p_\tau\,.
\ee
We have absorbed the coupling $g$ into redefinitions of the metric fields. This coupling can be recovered straightforwardly at the end of the calculation in the quantum effective action. In the rest of the paper, we will assume that both $G_{IJ}$ and $H_{IJ}$ are non-degenerate in the rest of the paper. 

In the background field method (see, \emph{e.g.}, \cite{Honerkamp:1971sh, Alvarez-Gaume:1981exa, Howe:1986vm, Callan:1989nz}), we split the worldvolume field $Y^I$ into a classical part $Y_0^I$ and a quantum fluctuation part, where the latter will be integrated out in the path integral that we will define in the next subsection. In the following, we construct a parametrization of this quantum fluctuation that is covariant with respect to the background geometry. Consider a sufficiently small neighborhood around $Y_0^I$\,. We introduce the interpolating function $Y^I_s$ of the affine parameter $s \in [0,1]$\,, such that
\be
	Y^I_{s=0} = Y^I_0\,,
		\qquad%
	Y^I_{s=1} = Y^I\,.
\ee 
We further require that the interpolating function $Y_s^I$ satisfy the geodesic equation,
\be \label{eq:geopi}
	\frac{d^2 Y^I_s}{ds^2} + \Pi^I{}_{KL} \, \frac{dY^K_s}{ds} \frac{dY^L_s}{ds} = 0\,,
\ee
where we introduced an abstract connection $\Pi$\,. We will fix $\Pi$ later in Section~\ref{sec:quaternion}. 
Consider a covariant quantum fluctuation $\ell^I$ that is defined to be a tangent vector along the geodesic,
\be \label{eq:defell}
	\ell^I = \frac{dY^I_s}{ds} \bigg|_{s=0}\,.
\ee
Define the derivatives covariantized with respect to the curved target space via
\be
	\CD_{\alpha} \, \ell^I = D_\alpha \, \ell^I + \Gamma^I{}^{}_{JK} \, D_\alpha Y^J \, \ell^K,
		\qquad%
	\Delta^{}_i \, \ell^I = \p^{}_i \, \ell^I + \Theta^I{}^{}_{JK} \, \p^{}_i Y^J \, \ell^K.
\ee
where the Christoffel symbols $\Gamma^I{}_{JK}$ and $\Theta^I{}_{JK}$ are defined in Eq.~\eqref{eq:csg}. 
Expanding the Lagrangian associated with the action \eqref{eq:se} with respect to $\ell^I$\,, we find
\be \label{eq:la}
	\CL = \CL^{(0)} + \CE^{}_K \, \ell^K + \frac{1}{2} \, \CO^{}_{KL} \, \ell^K \, \ell^L + O(\ell^3)\,, 
\ee
where, up to total derivatives,
\begin{subequations} \label{eq:s2}
\begin{align}
	\CL^{(0)} & = \frac{1}{2} \, \Bigl( \bar{D}^\alpha Y^I_0 \, D_\alpha Y^J_0 \, G^{}_{IJ} + 2 \, \p^{}_i Y^I_0 \, \p^{}_i Y^J_0 \, H^{}_{IJ} \Bigr)\,, \\[4pt]
	\CE^{}_K & = - \Bigl( \bar{\CD}^{\alpha} D_\alpha Y_0^I \, G^{}_{IK} + 2 \, \Box Y_0^I \, H^{}_{IK} \Bigr) \,, \label{eq:l1} \\[6pt]
	\CO^{}_{KL} & =  - \, G^{}_{KL} \, \bar{\CD}^\alpha \, \CD_{\!\alpha} - 2 \, H^{}_{KL} \, \Box + V^{}_{KL}\,. \label{eq:defco}
\end{align}
\end{subequations}
Recall that $\Box = \Delta_{i} \, \Delta_i$\,. 
We defined
\begin{align} \label{eq:ovb}
\begin{split}
	V_{KL} & =  \Bigl[ \bar{D}^\alpha Y_0^I \, D_\alpha Y_0^J \, R^M{}^{}_{KLJ} + \bar{\CD}^{\alpha} D_\alpha Y_0^I \, \bigl( \Pi^M{}^{}_{KL} - \Gamma^M{}^{}_{KL} \bigr) \Bigr] \, G^{}_{MI} \\[4pt]
		& \quad\quad\,\, + 2 \, \Bigl[ \p^{}_i Y_0^I \, \p^{}_i Y_0^J \, \Sigma^M{}^{}_{KLJ} + \Box Y_0^I \, \bigl( \Pi^M{}^{}_{KL} - \Theta^M{}^{}_{KL} \bigr) \Bigr] \, H^{}_{MI}\,.
\end{split}
\end{align}
Here, $R^I{}_{JKL}$ and $\Sigma^I{}_{JKL}$ in Eq.~\eqref{eq:rsigma} are the Riemann tensors associated with $G_{IJ}$ and $H_{IJ}$\,, respectively. The quantities $\Pi - \Gamma$ and $\Pi - \Theta$ denote the discrepancies between different Christoffel symbols and are both (1,\,2)-tensors. Here, all the background fields in Eq.~\eqref{eq:la} are understood to depend on $Y^I_0$\,. When we later integrate out the quantum fluctuation $\ell^I$ in the path integral in Section~\ref{sec:oleas}, the operator $\CO_{KL}$ in Eq.~\eqref{eq:defco} will be used to define a heat kernel equation whose solution computes the one-loop quantum corrections. 

\subsubsection{The classical propagator} \label{eq:clprop}

Due to the presence of the bimetric fields in the quadratic operator $\CO_{KL}$ in Eq.~\eqref{eq:defco}, the associated propagator does not have a simple form: Although it is possible to simultaneously diagonalize the metric fields, generically, there is \emph{no} coordinate transformation such that both $G^{}_{IJ}$ and $H^{}_{IJ}$ are flat locally at the same point on the target-space manifold. To demonstrate this explicitly, we perform a change of variable from $\ell^I$ to $\ell^A = E^{}_I{}^A \, \ell^I$\,, where $E^{}_I{}^A$\,, $A = 1, \cdots, d$ is the vielbein field that satisfies
\be
	G^{}_{IJ} = E^{}_I{}^A \, E^{}_J{}^B \, \delta^{}_{\!AB}\,.
\ee
We also define the inverse vielbein $E^I{}_A$ via the invertibility conditions,
\be
	E^I{}^{}_{\!A} \, E^{}_I{}^B = \delta_A^B\,,
		\qquad%
	E^I{}^{}_{\!A} \, E^{}_J{}^A = \delta_J^I\,.
\ee
It is possible to choose an $E^{}_I{}^A$ such that, locally, $H^{}_{IJ}$ is diagonalized, \emph{i.e.},
\be
	E^I{}^{}_{\!A} \, H^{}_{IJ} \, E^J{}^{}_{\!B} = \sum_A \lambda^{}_A \, \delta^{}_{AB}\,.
\ee 
Then, the Gaussian part in the Lagrangian \eqref{eq:la} is given by
\be
	\CL = - \frac{1}{2} \, \sum_{A} \ell^A \lr \bar{D}^\alpha D_\alpha + 2 \, \lambda^{}_A \, \p^{}_i \, \p^{}_i \, \rr \ell^A\,. 
\ee
Performing the Fourier transform with respect to the worldvolume coordinates $(t\,, x^i)$\,, and in terms of the frequency $\omega$ and spatial momentum $k^{}_i$\,, we find that the propagator is\,\footnote{The theory becomes regular in the IR when the relevant coupling $c$ or $m$ is turned on. In particular, it is interesting to emphasize that the theory is already IR finite when $c \neq 0$ but $m = 0$\,. Later in Section~\ref{sec:oleas}, we will introduce the IR regulator in a different way, which allows us to keep both $c$ and $m$ to be zero. The choice of IR regulators does not affect the resulting beta-functions, which only depend on the high energy behavior of the system around the $z=2$ Lifshitz point.}
\be
	\mathscr{G}^{AB} (\omega\,, \mathbf{k}\,; \theta\,, \theta') = - \frac{1}{4} \frac{\bar{D}^\alpha D_\alpha + 2 \, \lambda^{}_{\!A} \, k^2}{\omega^2 + \lambda_{\!A}^2 \, k^4} \, \delta^{AB} \, \delta^{(2)} (\theta - \theta')\,.
\ee 
Here, $k^2 = k^i \, k_i$\,. Note that \emph{no} sum is taken over the index $A$\,. The explicit dependence on $\lambda_A$ will generically lead to expressions that are difficult to be covariantized in the quantum effective action.
This brings interesting challenges to the evaluation of the beta-functions of Lifshitz sigma models. In Section~\ref{sec:oleas}, we will generalize the covariant heat kernel method in \cite{Yan:2021xnp} to incorporate supersymmetry, and then apply this method to compute the one-loop effective action for the Lifshitz sigma model defined by the action \eqref{eq:se}.

\subsubsection{Quaternions and bimetric symmetry} \label{sec:quaternion}

Before we proceed to evaluating the quantum action, a technical question has yet to be addressed: how to choose the connection $\Pi$ in the geodesic equation \eqref{eq:geopi}, with respect to which the quantum fluctuation $\ell$ in Eq.~\eqref{eq:defell} is defined on a local patch of the target-space manifold?

In the background field method, it is conventional to choose the background configuration $Y^I_0$ such that the Lagrangian term linear in $\ell$\,, \emph{i.e.}, $\CE_K$ in Eq.~\eqref{eq:l1}, vanishes. This gives rise to the equation of motion,
\be \label{eq:gheqn}
	 \bar{\CD}^{\alpha} D_\alpha Y_0^I \, G_{IK} (Y_0) + 2 \, \Box Y_0^I \, H_{IK} (Y_0) = 0 \,.
\ee
In general, implementing Eq.~\eqref{eq:gheqn} in the path integral may give rise to gravitational anomalies on the worldvolume, which should be absorbed into an appropriate worldvolume gravitational counterterm \cite{Polchinski:1998rq}. This does not concern us yet as we are focusing on the sigma model on flat worldvolume. However, without the inclusion of the worldvolume gravitational sector, our quantum effective action will only be defined up to a term proportional to the LHS of the equation of motion \eqref{eq:gheqn}. Such an arbitrariness is parametrized by the variable $\Pi^I{}_{JK}$\,, which is precisely the $\Pi$-dependence in the operator $\CO$ in Eq.~\eqref{eq:defco}. This $\Pi$-dependence is shown explicitly in Eq.~\eqref{eq:ovb} and is proportional to the LHS of Eq.~\eqref{eq:gheqn}.\,\footnote{Note that the same ambiguity also exists in string theory when the beta-functionals for the background metric field $G_{\mu\nu}$ and dilaton field $\Phi$ are considered individually. However, this ambiguity disappears in beta-functional of the T-dual invariant combination $\Phi - \frac{1}{4} \ln \det G_{\mu\nu}$\,.} 

Na\"{i}vely, it seems that the choice of $\Pi^K{}_{IJ}$ is arbitrary in the absence of any gravitational effects on the worldvolume. 
However, there are two important constraints. First, geometrically, for the geodesic equation \eqref{eq:geopi} to hold under the diffeomorphisms, the Christoffel symbol $\Pi^I{}_{JK}$ has to transform as  
\be
	\tilde{\Pi}^{I}{}_{JK} = \frac{\p \tilde{y}^{I}}{\p y^L} \frac{\p y^M}{\p \tilde{y}^{J}} \frac{\p y^N}{\p \tilde{y}^{K}} \, \Pi^M{}_{LN} + \frac{\p \tilde{y}^{I}}{\p y^L} \frac{\p^2 y^L}{\p \tilde{y}^{J} \, \p \tilde{y}^{K}}\,.
\ee
To match this transformation, a simple ansatz for $\Pi^I{}_{JK}$ is 
\be \label{eq:piy}
	\Pi^I{}_{JK} = v \, \Gamma^I{}_{JK} + (1-v) \, \Theta^I{}_{JK}\,,
\ee
where $\Gamma^I{}_{JK}$ and $\Theta^I{}_{JK}$ are defined in Eq.~\eqref{eq:csg} and $v$ is a real number.
In the following, we will present a second constraint that states Eq.~\eqref{eq:piy} has to be invariant under exchanging $G_{IJ}$ and $H_{IJ}$\,. 

We start with introducing a quaternion vector space over real numbers $a,\, b,\, c,\, d$\,, with the elements taking the form 
\be
	a + i \, b + j \, c + k \, d\,,
\ee
where the basis elements $i\,, j\,, k$ do not commute and satisfy
\be
	i^2 = j^2 = k^2 = i \, j \, k = -1\,.
\ee
Note that $a + b \, i$ defines a complex number. Then, we rewrite Eq.~\eqref{eq:se} as
 \be \label{eq:ser}
	S^{}_\text{E} = \frac{i}{g^2} \int d\tau \, d^2 \mathbf{x} \, d^2 \theta \, \Bigl[ D_- Y^I \, D_+ Y^J \, G_{IJ} + \bigl( k \, \bar{\p} Y^I \bigr) \, \bigl( j \, \p Y^J \bigr) \, H_{IJ} \Bigr]\,,
\ee
where $\p = \p_1 + i \, \p_2$ and $\bar{\p} = \p_1 - i \, \p_2$\,. The action \eqref{eq:ser} is unchanged under the replacements,
\be \label{eq:dtrnsf}
	G \longleftrightarrow H\,,
		\qquad%
	D_- \longleftrightarrow k \, \bar{\p}\,,
		\qquad%
	D_+ \longleftrightarrow j \, \p\,,
\ee
under which the anti-commutators $\{ D_-\,, D_+ \} = 0$ and $\{ k \, \bar{\p}, j \, \p \} = 0$ are mapped to each other. However, unlike $\{ D_\pm\,, D_\pm \} = 0$\,, the anti-commutators $\{ j \, \p, j \, \p \}$ and $\{ k \, \bar{\p}, k \, \bar{\p} \}$ do not vanish, but their appearance in any marginal (and relevant) operator in Eq.~\eqref{eq:ser} is forbidden due to spatial rotation symmetry.

Requiring that the expanded Lagrangian \eqref{eq:la} be invariant under the mapping \eqref{eq:dtrnsf} imposes the condition $v = 1/2$ in Eq.~\eqref{eq:piy}, such that the Christoffel $\Pi^I{}_{JK}$ is fixed to be
\be \label{eq:pgt}
	\Pi^I{}_{JK} = \frac{1}{2} \Bigl( \Gamma^I{}_{JK} + \Theta^I{}_{JK} \Bigr)\,.
\ee
The geodesic equation \eqref{eq:geopi} becomes
\be \label{eq:gezy}
	\frac{d^2 Y^I_s}{d s^2} + \frac{1}{2} \Bigl[ \Gamma^I{}_{KL}(Y_s) + \Theta^I{}_{KL}(Y_s) \Bigr] \, \frac{dY^K_s}{ds} \, \frac{dY^L_s}{ds} = 0\,.
\ee
Note that Eqs.~\eqref{eq:pgt} and \eqref{eq:gezy} remain invariant after swapping $G_{IJ}$ and $H_{IJ}$\,.
As a result, the potential term $V_{KL}$ in Eq.~\eqref{eq:ovb} becomes
\begin{align} \label{eq:vfixed}
\begin{split}
	V_{KL} & =  \Bigl[ \bar{D}^\alpha Y_0^I \, D_\alpha Y_0^J \, R^M{}^{}_{KLJ} - \tfrac{1}{2} \, \bar{\CD}^{\alpha} D_\alpha Y_0^I \, S^M{}^{}_{KL} \Bigr] \, G^{}_{MI} \\[4pt]
		& \quad\quad\,\,\, + 2 \, \Bigl[ \p^{}_i Y_0^I \, \p^{}_i Y_0^J \, \Sigma^M{}^{}_{KLJ} + \tfrac{1}{2} \, \Box Y_0^I \, S^M{}^{}_{KL} \Bigr] \, H^{}_{MI}\,.
\end{split}
\end{align}
Plugging the expression \eqref{eq:vfixed} into the operator $\CO_{IJ}$ in Eq.~\eqref{eq:defco} yields the final operator that we will use to compute the one-loop effective action in Section~\ref{sec:oleas}. 

The invariance of the action \eqref{eq:ser} under the replacements in Eq.~\eqref{eq:dtrnsf} is reminiscent of the case for two-dimensional Lifshitz sigma models coupled to a bimetric target-space geometry, which is described by the action \cite{Yan:2021xnp}
\be \label{eq:sact}
	S_\text{string} = \frac{1}{4\pi\alpha'} \int d\tau \, dx \, \Bigl[ \p^{}_\tau X^I \, \p^{}_\tau X^J \, G^{}_{IJ} (X) + \p^{}_x X^I \, \p^{}_x X^J \, H^{}_{IJ} (X) \Bigr]\,.
\ee
The sigma model \eqref{eq:sact} is invariant under the replacement,
\be \label{eq:replacementrules2d}
	G \longleftrightarrow H\,, 
		\qquad%
	\tau \longleftrightarrow x\,,
\ee
where the second rule implies that we swap $\p_\tau$ and $\p_x$\,.
In the two-dimensional Lifshitz sigma model \eqref{eq:sact}, the beta-functionals of the metric fields are mapped to each other upon swapping $G_{IJ}$ and $H_{IJ}$ \cite{Yan:2021xnp}. However, there is no analog of Eq.~\eqref{eq:replacementrules2d} for the three-dimensional sigma model \eqref{eq:ser}, where the derivatives $D_\alpha$ and $\p_i$ have distinct structures. This implies that the beta-functionals for $G_{IJ}$ and $H_{IJ}$ in the three-dimensional theory define by Eq.~\eqref{eq:se} are expected to have different structures.

\subsection{One-Loop Effective Action in Superspace} \label{sec:oleas}

We now develop the general theory of supercovariant heat kernel method, which we will then apply to the operator $\CO_{KL}$ in Eq.~\eqref{eq:defco} to compute the one-loop quantum effective action for the theory defined by Eq.~\eqref{eq:se}. The following procedure directly generalizes the method proposed in \cite{Yan:2021xnp} for evaluating the one-loop beta-functions of Lifshitz sigma models describing strings propagating in a bimetric spacetime. Also see \cite{Gusynin:1989ky, Grosvenor:2021zvq} for closely related discussions on the heat kernel method.

We start with defining the path integral associated with the action \eqref{eq:se}, for which a choice of the reference metric on the functional space of the quantum fluctuation $\ell^I$ has to be made. Since the sigma model is defined on a bimetric spacetime, the metric fields $G_{IJ}$ and $H_{IJ}$ are equally natural candidates for the reference metric. For example, if $G_{IJ}$ is the reference metric, the associated path integral with a source $J_I$ is
\be \label{eq:pathintegral}
	\CZ = \int d\ell^I \sqrt{-G(Y_0)} \, \exp \lc -\hbar \ls S_\text{E} (Y_0\,, \, \ell) - \int d\tau\, d^2 \mathbf{x} \, d^2 \theta \, J^{}_I \, \ell^I \rs \rc,
\ee
where $G(Y_0)$ is evaluated at the background field value $Y_0^I$\,. Regardless of the choice of the reference metric, the final effective action is uniquely defined. We will derive the one-loop effective actions using both of the reference metrics $G_{IJ}$ and $H_{IJ}$ and show that they corroborate each other. This will serve as a powerful check of the RG calculation. We also choose the value of $Y^I_0$ in Eq.~\eqref{eq:l1} such that $\CE^{}_I = J^{}_I$\,. This procedure eliminates any terms that are linear in $\ell^I$\,.

\subsubsection{The heat kernel equation}

Integrating out $\ell^I$ in the path integral \eqref{eq:pathintegral}, we find in the semi-classical limit,
\be
	\CZ \sim \exp \Bigl[ - S^{(0)} - \hbar \, \Gamma_\text{1-loop} + O(\hbar^2) \Bigr]\,,
\ee
with the one-loop quantum effective action defined by
\be \label{eq:goloop}
	\Gamma_\text{1-loop} = \frac{1}{2} \, \tr \log \Bigl( \CO_{IK} \, G^{KJ} / \mu_\text{IR}^{2} \Bigr)\,,
\ee
where $\mu^{}_\text{IR}$ is an IR regulator.
The one-loop effective action can be further rewritten as
\be \label{eq:gol}
	\Gamma_\text{1-loop} = - \frac{1}{2} \frac{d}{ds} \bigg|_{s=0} \frac{\mu_\text{IR}^{2s}}{\Gamma(s)} \int d\tau \, d^2\mathbf{x} \, d^2 \theta\int^\infty_{0} d o \, o^{s-1} \, \CK^{}_I{}^I (\phi\,, \phi \, | \, o)\,,
\ee
where we defined the collective coordinate $\phi \equiv (t, \mathbf{x}, \theta^\alpha)$\,. 
We also introduced the kernel, 
\be \label{eq:ckdef0}
	\CK = \exp \Bigl( -o \, \CO \, G^{-1} \Bigr)\,,
\ee 
which satisfies the heat kernel equation,
\be \label{eq:hkeqn}
	\Bigl( \delta_I^L \, \p_o + \CO^{}_{IK} \, G^{KL} \Bigr) \, \CK^{}_L{}^J \bigl(\phi\,, \phi^{}_0 \, | \, o \bigr) = 0\,.
\ee
The general solution to the heat kernel equation \eqref{eq:hkeqn} can be expressed in terms of the following integral form using the resolvent $\CG$ of the operator $\CO \, G^{-1}$\,:
\be \label{eq:kij}
	\CK^{}_I{}^J (\phi, \phi^{}_0 | o) = \int_\CC \frac{i \, d\lambda}{2\pi} \, e^{-o \, \lambda} \, \CG^{}_I{}^J (\phi, \phi^{}_0 | \lambda)\,,
\ee
where $\CC$ is a contour that bounds the spectrum of the operator $\CO$ in the complex plane and is traversed in the counter-clockwise direction. The resolvent
\be \label{eq:defGprop}
	\CG (\phi, \phi^{}_0 | o) = \bigl\langle \phi \big| \, \bigl( \CO \, G^{-1} - \lambda \, \mathbb{1} \bigr)^{-1} \, \big| \phi^{}_0 \bigr\rangle
\ee
acts as a ``propagator." 

\subsubsection{Fourier transform in superspace}

To proceed in the heat kernel method, we will need to perform a Fourier transform of $\CG^{}_I{}^J$\,. For this purpose, we first define the following Fourier transform of an arbitrary function $U(\theta)$ of the Grassmannian coordinate $\theta$\,:
\be \label{eq:gft}
	\tilde{U} (\xi) = \int \frac{d^2 \theta}{(i/2)^2} \, e^{-\bar{\theta} \, \xi} \, U(\theta)\,.
\ee
Here, $\xi^{}_\alpha$ denotes the Grassmannian momentum conjugate to $\theta^\alpha$.
Expand $U (\theta)$ with respect to $\theta$ as
\be
	U (\theta) = u + \bar{\theta} \, \gamma + \frac{1}{2} \, \theta^2 \, v\,,
\ee 
we find that its Fourier transform is
\be
	\tilde{U} (\xi) = - 2 \, v - 2 \, \bar{\xi} \, \gamma + u \, \xi^2\,.
\ee
It is then straightforward to show that the inverse Fourier transform gives
\be \label{eq:igft}
	U(\theta) = \int d^2 \xi \, e^{\bar{\xi} \, \theta} \, \tilde{U} (\xi) \,.
\ee
For example, when $U (\theta) = 1$\,, we find the Fourier transform \eqref{eq:gft} and its inverse \eqref{eq:igft} give rise to the following expected identities, respectively:
\be
	\delta^{(2)} (\xi) = \int \frac{d^2 \theta}{(i/2)^2} \, e^{- \bar{\theta} \, \xi}\,,
		\qquad%
	1 = \int d^2 \xi \, \delta^{(2)} (\xi)\,.
\ee

Using the above Grassmannian generalization of the Fourier transform, we write $\CG$ in Eq.~\eqref{eq:defGprop} as
\begin{align} \label{eq:osigmaeqn}
\begin{split}
	\CG^{}_I{}^J (\phi\,, \phi^{}_0 | o) = \!\! \int \! \frac{d\omega}{2\pi} \, \frac{d^2\mathbf{k}}{(2\pi)^2} \, \frac{d^2 \xi}{(i/2)^2} \, & \exp \Bigl[ i \, \omega \, (\tau - \tau_0) + i \, \mathbf{k} \! \cdot \! (\mathbf{x} - \mathbf{x}_0) + i \, \xi^\intercal \, \rho^0 \, (\theta - \theta_0) \Bigr] \\[2pt]
	& \hspace{3.3cm} \times \sigma^{}_I{}^J \Bigl( \phi\,, \phi^{}_0\,, \{ \omega\,, \mathbf{k}\,, \xi \} \Big| \lambda \Bigr)\,,
\end{split}
\end{align}
where $\sigma$ satisfies
\be
	\Bigl[ \CO^{}_{IK} \bigl( \phi\,; \CD_\alpha + i \, \Xi_\alpha\,; \Delta_i + i \, k_i \bigr) - \lambda \, G^{}_{IK} (\phi) \Bigr] \, \sigma^{KJ} \Bigl( \phi\,; \phi_0\,, \{ \Xi\,, \mathbf{k} \} \Big| \lambda \Bigr) = I^{}_I{}^J (\phi\,; \phi_0)\,.
\ee
Here, 
\be \label{eq:defq}
	\Xi_\alpha = - \xi_\alpha + i \, \omega \, \frac{\p}{\p \xi^\alpha}\,.
\ee
Moreover, $\sigma^{IJ} = G^{IK} \, \sigma^{}_K{}^J$ and $I^{}_I{}^J$ is in general a bi-function that satisfies certain conditions in the coincidence limit with $\phi \rightarrow \phi_0$ \cite{Gusynin:1989ky, Grosvenor:2021zvq}. For our purpose, it is sufficient to consider the simplified equation,
\be \label{eq:sigmaeqn}
	\Bigl[ \CO^{}_{IK} (\phi\,; \CD_\alpha + i \, \Xi_\alpha\,, \Delta_i + i \, k_i) - \lambda \, G_{IK} (\phi) \Bigr] \, \sigma^{KJ} \Bigl(\phi\,, \{ \Xi\,, \mathbf{k} \} \Big| \lambda \Bigr) = \delta_I^J\,.
\ee
In the following, we relate the quantity $\sigma^{IJ}$ to the one-loop effective action by introducing a supersymmetric generalization of the Seeley-Gilkey asymptotic expansion around $o \rightarrow 0^+$ \cite{seeley1967complex, Gilkey:1975iq}.

\subsubsection{Seeley-Gilkey coefficients} \label{sec:sgcea}

The traced heat kernel $\CK$ from the definition \eqref{eq:ckdef0} has the following asymptotic expansion around $o \rightarrow 0^+$\,:
\be \label{eq:aseck}
	\CK^{}_I{}^I (\phi, \phi | o) = \sum_{m=0}^\infty E_m (\phi) \, o^{\frac{m - (z + D - D^{}_{\!\theta})}{\Delta}}\,,
\ee
where $\Delta = 2$ is the degree of the operator $\CO$ in Eq.~\eqref{eq:defco}, measured in spatial momentum, $z = 2$ is the dynamical critical exponent that measures the space and time anisotropy of $\CO$\,, $D = 2$ is the spatial dimension of the (2+1)-dimensional manifold on which the sigma model is defined, and $D^{}_\theta = 2$ counts the number of the Grassmannian directions labeled by $\theta_\alpha$\,. Consequently, Eq.~\eqref{eq:aseck} becomes
\be \label{eq:asecksiim}
	\CK_I{}^I (\phi, \phi | o) = \sum_{m=0}^\infty E_m (\phi) \, o^{\frac{m - 2}{2}}\,.
\ee
We also expand $\sigma$ as
\be \label{eq:exsi}
	\sigma = \sum_{m=0}^\infty \sigma_m\,,
		\qquad%
	\sigma_m \Bigl(\phi\,, \{ b \, q\,, b \, \mathbf{k} \} \, \Big| \, b^2 \, \lambda \Bigr) = b^{-m-\Delta} \, \sigma_m \Bigl(\phi\,, \{ \Xi\,, \mathbf{k} \} \, \Big| \, \lambda\Bigr)\,,
\ee
with $\Delta = 2$\,.
Note that both $\Xi$ and $\mathbf{k}$ are rescaled by the momentum factor $b$\,. Implicitly, the frequency $\omega$ that enters via Eq.~\eqref{eq:defq}. These rescalings are consistent with the scaling dimensions $[\omega] = 1$ and $[\xi] = [\mathbf{k}] =1/2$\,.
The expressions of $\sigma^{}_m$ can be solved recursively in order of $b$ by plugging Eq.~\eqref{eq:exsi} into Eq.~\eqref{eq:sigmaeqn}, 
and the solutions are related to the Seeley-Gilkey coefficients $E_m$ via
\be \label{eq:emphi}
	E^{}_m (\phi) = \int \frac{d\omega}{2\pi} \frac{d^2\mathbf{k}}{(2\pi)^2} \frac{d^2 \xi}{(i/2)^2} \oint_\CC \frac{i \, d\lambda}{2\pi} \, e^{-\lambda} \, G^{}_{IJ} (\phi) \, \sigma_m^{IJ} \Bigl(\phi\,, \{ \Xi\,, \mathbf{k}\} \Big| \lambda\Bigr)\,,
\ee
which is derived by plugging Eqs.~\eqref{eq:osigmaeqn}, \eqref{eq:aseck} and \eqref{eq:exsi} into the trace of Eq.~\eqref{eq:kij} followed by setting $b = o^{-1/2}$\,.
Finally, plugging Eq.~\eqref{eq:asecksiim} into Eq.~\eqref{eq:gol}, we find that the one-loop effective action is related to $E_m$ in Eq.~\eqref{eq:emphi} via
\begin{align} \label{eq:olea}
\begin{split}
	\Gamma_\text{1-loop} & = - \frac{1}{2} \frac{d}{ds} \bigg|_{s=0} \frac{\mu^{2s}_{\text{IR}}}{\Gamma(s)} \int d\tau \, d^2 \mathbf{x} \, d^2 \theta \int_{1/\Lambda^2}^\infty do \, o^{s-1} \, \CK^{}_I{}^I (\phi\,, \phi | o) \\[2pt]
		& = - \frac{1}{2} \int d\tau \, d^2\mathbf{x} \, d^2\theta \, \left\{ E_0 \, \Lambda^2 + 2 \, E_1 \, \Lambda + E_2 \, \log \lr \frac{\Lambda^2}{\mu^2_\text{IR}} \rr + \text{finite} \right\}\,.
\end{split}
\end{align}
Note that we have regularized the lower bound of the integral over the heat kernel parameter $o$\,, with $\Lambda$ acting as a UV energy sharp cutoff. 
Imposing appropriate renormalization conditions and fine tuning the counterterms such that all the power-law divergences are cancelled, we obtain
\begin{align} \label{eq:olea2}
\begin{split}
	\Gamma_\text{1-loop} & = - \frac{1}{2} \int d\tau \, d^2\mathbf{x} \, d^2\theta \, E_2 \, \ln \! \lr \frac{M^2}{\mu^2_\text{IR}} \rr,
\end{split}
\end{align}
where $M$ is the renormalization scale. The associated beta-functions can be read off from the above one-loop effective action.

\subsubsection{Solving the recursion relations} \label{sec:srr}

We are now ready to solve for the Seeley-Gilkey coefficients $E_m$\,. Plugging Eq.~\eqref{eq:defco} into Eq.~\eqref{eq:sigmaeqn}, followed by applying Eq.~\eqref{eq:exsi}, we find the following recursion relations:
\begin{subequations} \label{eq:rrs}
\begin{align}
	\CA^{}_{IK} \,\sigma_0^{KJ} & = \delta_I^J, \\[4pt]
	\CA^{}_{IK} \, \sigma_1^{KJ} - 2 \, i \lr G^{}_{IK} \, \overline{\Xi}^{\,\alpha} \CD_{\alpha} + 2 \, H_{IK} \, k^i \Delta_i \rr \sigma_0^{KJ} & = 0\,, \\[4pt]
	\CA^{}_{IK} \, \sigma_2^{KJ} - 2 \, i \lr G^{}_{IK} \, \overline{\Xi}^{\,\alpha} \CD^{}_{\alpha} + 2 \, H^{}_{IK} \, k^i \Delta_i \rr \sigma_1^{KJ} & \notag \\[4pt] 
	- \lr G^{}_{IK} \, \bar{\CD}^\alpha \CD_\alpha + 2 \, H^{}_{IK} \, \Delta^i \Delta_i - V^{}_{IK} \rr \sigma_0^{KJ} & = 0\,,
\end{align}
\end{subequations}
where
\be \label{eq:CA}
	\CA^{}_{IJ} = \Omega \, G^{}_{IJ} + 2 \, \Phi \, H^{}_{IJ}\,,
		\qquad%
	\Omega = \Xi^2 - \lambda\,,
		\qquad%
	\Phi = k^2\,,
\ee
and $\Xi_\alpha$ has been defined in Eq.~\eqref{eq:defq}. It is convenient to define the matrices
\begin{align} \label{eq:ckdef}
	\CK 
	= \lr k^2 - \frac{\lambda}{2} \rr \mathbb{1} - \Phi \, f \,,
		\qquad%
	f = \mathbb{1} - \CH\,,
		\qquad%
	\CH = G^{-1} \, H\,.
\end{align}
We will drop the unit matrix ``$\mathbb{1}$" in the following. The inverse of the operator $\CA_{IJ}$ is given by $\bigl( \Gamma \, G^{-1} \bigr){}^{IJ}$\,, where $\Gamma^I{}_J$ is defined to be
\be \label{eq:gammaprop}
	\Gamma = \frac{2 \, \CK - \Xi^2}{4 \, \bigl( \omega^2 + \CK^2 \bigr)}\,,
\ee
which satisfies the operator equations $\CA \, \bigl( \Gamma \, G^{-1} \bigr) = \bigl( \Gamma \, G^{-1} \bigr) \CA = \mathbb{1}$\,. We are ultimately interested in the Seeley-Gilkey coefficient $E^{}_2$ in Eq.~\eqref{eq:emphi}, where integrals over the momenta $\omega$\,, $\mathbf{k}$ and $\xi_\alpha$ are performed. Therefore, we will drop all terms in $\sigma^{}_2$ that are odd in these momenta, as they have no contribution to $E^{}_2$\,. We find
\begin{subequations} \label{eq:solvrecur}
\begin{align}
	\sigma^{}_0 & = \Gamma \, \mathbb{1}\,, 
		&%
	D^{}_\alpha \sigma^{}_0 & = - 2 \, \Phi \, \Gamma \, H_\alpha \, \sigma^{}_0\,, \\[4pt]
	\sigma^{}_1 & = 2 \, i \, \Gamma \, \bigl( \, \overline{\Xi}^\alpha \, \CD^{}_\alpha + 2 \, H \, k^i \Delta^{}_i \bigr) \sigma^{}_0\,,
		&%
	\Delta^{}_i \sigma^{}_0 &= - \Gamma \, \Omega \, G^{}_i \, \sigma^{}_0\,, \\[4pt]
	\CD^{}_\alpha \CD^{}_\beta \sigma^{}_0 & = - 2 \, \Phi \, \Gamma \bigl( \, H^{}_{\alpha\beta} + 2 \, H^{}_{[\alpha} \, \CD^{}_{\beta]} \bigr) \, \sigma^{}_0\,, 
		&%
	\CD^{}_- \sigma^{}_1 &\sim \Gamma \bigl( \, \Xi^{}_- \, \CD^{}_- \CD^{}_+ \sigma^{}_0 - 2 \, \Phi \, H^{}_- \, \sigma^{}_1 \bigr)\,, \\[4pt]
	\Delta^{}_i \Delta^{}_j \sigma^{}_0 & = - \Gamma \, \Omega \, \bigl( \, G^{}_{ij} + 2 \, G^{}_{(i} \, \Delta^{}_{j)} \bigr) \, \sigma^{}_0\,, 
		&%
	\CD^{}_+ \sigma^{}_1 &\sim \Gamma \bigl(  - \Xi^{}_+ \, \CD^{}_+ \CD^{}_- \sigma^{}_0 - 2 \, \Phi \, H^{}_+ \, \sigma^{}_1 \bigr)\,,  \\[2pt]
	& &
	\Delta^{}_i \sigma^{}_1 & \sim \Gamma \bigl( 4 \, i \, H \, k^j \, \Delta^{}_i \, \Delta^{}_j \, \sigma^{}_0 - \Omega \, G^{}_i \, \sigma^{}_1 \bigr)\,.
\end{align}
\end{subequations}
Here, $H^{}_{\alpha_1 \cdots \alpha_n} \equiv \CD^{}_{\alpha_1} \cdots \CD^{}_{\alpha_n} H$ and $G^{}_{i_1 \cdots i_n} = \Delta_{i_1} \cdots \Delta_{i_n} G$\,. Finally, the coefficient $\sigma_2$ can be calculated by using 
\begin{align}
\begin{split}
	\sigma_2 & = \Gamma \Bigl[ 2 \, i \bigl( \, \overline{\Xi}^\alpha \, \CD^{}_\alpha + 2 \, H \, k^i \Delta^{}_i \bigr) \sigma^{}_1 + \bigl( \bar{\CD}^\alpha \CD^{}_\alpha + 2 \, H \Delta^{\!i} \Delta^{}_i \bigr) \sigma^{}_0 - V \sigma^{}_0 \Bigr] \\[2pt]
		& \sim \bigl( 2 \, i \, \Gamma \, \overline{\Xi}^\alpha \CD^{}_\alpha \bigr)^{\!2} \sigma^{}_0 + \bigl( 4 \, i \, \Gamma \, H \, k^i \, \Delta^{}_i \bigr)^{\!2} \sigma^{}_0 + \Gamma \bigl[ \bigl( \bar{\CD}^\alpha \CD^{}_\alpha + 2 \, H \Delta^{\!i} \Delta^{}_i \bigr) \sigma^{}_0 - V \sigma^{}_0 \bigr]\,.
\end{split}
\end{align}
In the above expressions, ``$\sim$" implies that terms contributing trivially $E_2$ are dropped.
Using Eq.~\eqref{eq:emphi}, we find
\be \label{eq:e2gresult}
	E_2 = - \frac{1}{24 \, \pi} \, \tr \Bigl[ \Delta^{\!i} \bigl( G^{-1} \, f_i \, \bigr) \Bigr] + \frac{1}{16 \pi} \, \tr \! \ls \frac{1}{(\mathbb{1} - f)^2} \lr \frac{\bar{f}^\alpha \, f_\alpha}{\mathbb{2} - f - \tilde{f}} + \frac{1}{2} \, f^\alpha{}_\alpha \rr - \frac{2 \, \CV}{\mathbb{1}-f} \rs.
\ee
Here,
\begin{align}
	f_{\alpha_1 \cdots \alpha_n} & \equiv \CD_{\alpha_1} \! \cdots \CD_{\alpha_n} f\,,
		&%
	f_{i_1 \cdots i_n} & \equiv \Delta_{i_1} \! \cdots \Delta_{i_n} f\,,
		&%
	f_\alpha{}^\alpha &\equiv \bar{\CD}^\alpha \CD_\alpha f\,,
		&%
	\CV &\equiv G^{-1} \, V\,.
\end{align}
We also introduced the notation $\tilde{f}$ to imply that the related matrix $f$ is located between $\bar{f}^\alpha = \bar{D}^\alpha f$ and $f_\alpha$\,, \emph{i.e.}, 
$\tilde{f}^n \, \bar{f}^\alpha f_\alpha \equiv \bar{f}^\alpha f^n \, f_\alpha$ and, in Eq.~\eqref{eq:e2gresult},
\be
	\frac{\bar{f}^\alpha \, f_\alpha}{2 - f - \tilde{f}} = \sum_{n=0}^\infty \frac{1}{(2-f)^{n+1}} \, \bar{f}^\alpha \, f^{n} \, f_\alpha\,.
\ee
Expressing Eq.~\eqref{eq:e2gresult} in terms of the metric fields $H$ and $G$\,, we find
\begin{align} \label{eq:e2gh}
\begin{split}
	E_2 = \frac{1}{24 \, \pi} \, \tr \Bigl[ \Delta^{\!i} \bigl( G^{-1} \, G_i \, \bigr) \Bigr] & + \frac{1}{16\pi} \, \tr \biggl[ \CH^{-2} \, \bigl( \CH + \tilde{\CH} \bigr)^{-1} \, \CH^\alpha \, \CH_\alpha \biggr] \\[2pt]
	& - \frac{1}{32\pi} \, \tr \Bigl( \CH^{-2} \, {\CH}^\alpha{}_\alpha \Bigr) - \frac{1}{8\pi} \, \tr \Bigl( \CH^{-1} \, \CV \Bigr)\,,
\end{split}
\end{align}
where $\CH$ has been defined in Eq.~\eqref{eq:ckdef}. The notation $\tilde{\CH}$ indicates that the related matrix is located between $\CH^\alpha$ and $\CH_\alpha$\,. 
More explicitly, Eq.~\eqref{eq:e2gh} represents
\begin{align} \label{eq:E2G}
\begin{split}
	E_2 = - \frac{1}{24 \, \pi} \, \tr \Bigl[ \Delta^{\!i} \bigl( G^{-1} \Delta^{}_i G \, \bigr) \Bigr] & + \frac{1}{16\pi} \, \tr \biggl[ \CH^{-2} \, \sum_{n=0}^\infty \bigl( \mathbb{1} + \CH \bigr)^{-n-1} \, \bar{\CD}^\alpha \CH \, f^n \, \CD^{}_{\!\alpha} \CH \biggr] \\[2pt]
	& - \frac{1}{32\pi} \, \tr \Bigl( \CH^{-2} \, \bar{\CD}^\alpha \CD^{}_{\!\alpha} \CH \Bigr) - \frac{1}{8\pi} \, \tr \Bigl( \CH^{-1} \CV \Bigr)\,,
\end{split}
\end{align}
We also note the relation from integration by parts, 
\be \label{eq:ibp}
	\tr \Bigl( \CH^{-2} \, \bar{\CD}^\alpha \CD_\alpha \CH \Bigr) = \tr \Bigl[ \bar{\CD}^\alpha \! \lr \CH^{-2} \, \CD_\alpha \CH \rr + 2 \, \CH^{-2} \, \bar{\CD}^\alpha \CH \, \CH^{-1} \, \CD_\alpha \CH \Bigr]\,.
\ee
Finally, plugging Eqs.~\eqref{eq:e2gh} and \eqref{eq:ibp} into Eq.~\eqref{eq:olea2}, we obtain the following log-divergent contribution to the one-loop effective action:
\begin{align} \label{eq:1lefar}
	\Gamma_\text{1-loop} & = \frac{\ln \! \lr M / \mu_\text{IR} \rr}{16 \, \pi} \int d\tau \, d^2\mathbf{x} \, d^2\theta \, 
	 \tr \ls \CH^{-1} \, \bigl( \CH + \tilde{\CH} \bigr)^{-1}  \, \bar{\CD}^\alpha \CH \, \CH^{-1} \, \CD_\alpha \CH + 2 \, \CH^{-1} \, \CV \rs.
\end{align}
Here, assuming that appropriate boundary conditions are imposed, we have dropped the boundary terms. It is interesting to note that, as a consequence of supersymmetry, the contributions to terms quadratic in spatial derivatives other than the ones in $\CV$ defined in Eq.~\eqref{eq:ovb} vanish identically.

\subsubsection{Changing the reference metric}

When we define the path integral associated with the quadratic action in Eq.~\eqref{eq:pathintegral}, a choice of the background metric has to be made in order to define the measure over the target-space geometry. However, after integrating out the the quantum fluctuation in the path integral, the resulting one-loop effective action should not be affected by this choice of a reference metric. 
In Eq.~\eqref{eq:pathintegral}, the measure is defined using the reference metric $G_{IJ}$\,. In the present subsection, we will perform the same path integral but now with respect to $H_{IJ}$ as a reference metric, which will act as a nontrivial crosscheck of the resulting one-loop effective action \eqref{eq:1lefar}. 

We now define the path integral \eqref{eq:pathintegral} equivalently as
\be \label{eq:pathintegralh}
	\CZ = \int d\ell^I \sqrt{-H(Y_0)} \, e^{-S_\text{E} (Y)}\,,
\ee
with respect to the reference metric $H_{IJ}$ instead of $G_{IJ}$\,. 
The one-loop quantum effective action \eqref{eq:goloop} now becomes
\be
	\Gamma_\text{1-loop} = \frac{1}{2} \, \tr \ln \Bigl( \CO_{IK} \, H^{KJ} / \mu^2_\text{IR} \Bigr)\,.
\ee
Consequently, the Seeley-Gilkey coefficients $E_m$'s in \eqref{eq:emphi} is replaced with
\be 
	E_m (\phi) = \int \frac{d\omega}{2\pi} \frac{d^2\mathbf{k}}{(2\pi)^2} \frac{d^2 \xi}{(i/2)^2} \oint_\CC \frac{i \, d\lambda}{2\pi} \, e^{-\lambda} \, H_{IJ} (\phi) \, \sigma_m^{IJ} (\phi\,, \{ q, \mathbf{k}\} | \lambda)\,,
\ee
where $\sigma_m$ is still defined by Eq.~\eqref{eq:exsi}. The symbol $\sigma$ now satisfies
\be 
	\Bigl[ \CO_{IK} \bigl( \phi\,; \CD_\alpha + i \, \Xi_\alpha\,; \Delta_i + i \, k_i \bigr) - \lambda \, H_{IK} (\phi) \Bigr] \, \sigma^{KL} \Bigl(\phi\,, \{ \Xi\,, \mathbf{k} \} \Big| \lambda\Bigr) = \delta_I^J\,,
\ee
instead of Eq.~\eqref{eq:sigmaeqn}. 

The recursion relations are in form the same as Eq.~\eqref{eq:rrs}. However, the quantities $\Omega$ and $\Phi$ in Eq.~\eqref{eq:CA} have to be replaced with
\be
	\Omega = \Xi^2\,,
		\qquad%
	\Phi = k^2 - \frac{\lambda}{2}\,.
\ee
Moreover, $\CK$ in Eq.~\eqref{eq:ckdef} becomes
\begin{align} 
	\CK = \lr k^2 - \frac{\lambda}{2} \rr \, \CH \,.
\end{align}
The solutions to the recursion relations are also in form the same as \eqref{eq:solvrecur}, but with the above new definitions taken into account. Performing a similar analysis as in Section~\ref{sec:srr}, we find
\begin{align} \label{eq:e2h}
	E_2 
	= \frac{1}{16 \, \pi} \, \tr \ls \bigl( \CH + \tilde{\CH} \bigr)^{-1} \Bigl( \CH^{-2} + \tilde{\CH}^{-2} + \CH^{-1} \, \tilde{\CH}^{-1} \Bigr) \, \bar{\CD}^\alpha \CH \, \CD_\alpha \CH \rs & \notag \\[4pt]
	- \frac{1}{16 \, \pi} \, \tr \Bigl( \CH^{-2} \, \bar{\CD}^\alpha \CD_\alpha \CH \Bigr) - \frac{1}{8 \, \pi} \tr \Bigl( \CH^{-1} \, \CV \Bigr) &\,. 
\end{align} 
Using the integration-by-part relation \eqref{eq:ibp}, we find that Eq.~\eqref{eq:e2h} corroborates the one-loop effective action \eqref{eq:1lefar}.

\subsection{Renormalization Group Flows} \label{sec:rgf}

We have derived the one-loop effective action in Eq.~\eqref{eq:1lefar}. Together with the classical action \eqref{eq:se}, we find that the total quantum effective action is
\begin{align} 
\begin{split}
	S & = \frac{1}{2 \, g^2} \int d\tau \, d^2 \mathbf{x} \, d^2 \theta \, \Bigl[ \bar{D}^\alpha Y_0^I \, D_\alpha Y_0^J \, \bigl( G_{IJ} + \delta G_{IJ} \bigr) + 2 \, \p^i Y^I \, \p_i Y^J \, \bigl( H_{IJ} + \delta H_{IJ} \bigr) \Bigr]\,.
\end{split}
\end{align}
where
\begin{subequations}
\begin{align}
	\delta G_{IJ} = - \ln \! \lr \frac{M}{\mu_\text{IR}} \rr \beta^G_{IJ}\,, 
		\qquad%
	\delta H_{IJ} = - \ln \! \lr \frac{M}{\mu_\text{IR}} \rr \beta^H_{IJ}\,,
\end{align}
\end{subequations}
Here, the beta-functionals $\beta^G$ and $\beta^H$ are given by
\begin{subequations} \label{eq:betaGH}
\begin{align}
	\beta^G_{IJ} & = \frac{g^2}{4 \, \pi} \, \Bigl[ - H^{KL} \, R^M{}^{}_{KL(I}  \, G^{}_{J)M} - \tfrac{1}{2} \, G^{}_{M(I} \nabla^{}_{\!J)} \bigl( H^{KL} \, S^M{}^{}_{KL} \bigr) \Bigr] \notag \\[2pt]
	& \hspace{2.5cm} - \frac{g^2}{8 \, \pi} \, \tr \Bigl[ \CH^{-1} \, \bigl( \CH + \tilde{\CH} \bigr)^{-1} \, \nabla^{}_{\!(I} \CH \,\, \CH^{-1} \, \nabla^{}_{\!J)} \CH \Bigr] + O(g^4)\,, \\[4pt]
	\beta^H_{IJ} & = \frac{g^2}{4 \, \pi} \, \Bigl[ \Sigma^{}_{IJ} + \tfrac{1}{2} \, H^{KL} \, H^{}_{M(I} \, \Delta^{}_{J)} S^M{}^{}_{KL} \Bigr] + O(g^4)\,.
\end{align}
\end{subequations}
Note that we already plugged Eq.~\eqref{eq:vfixed} into $\CV = G^{-1} \, V$ and also recovered the dependence on the coupling $g$\,. We also defined the Ricci tensor $\Sigma_{IJ} = \Sigma^K{}_{I K J}$\,.
Recall that $\CH = G^{-1} \, H$ and the tilde in $\tilde{\CH}$ indicates that such a matrix is inserted between $\nabla^{}_{\!I} \CH$ and $\nabla^{}_{\!\!J} \CH$\,, \emph{i.e.},
\begin{align}
\begin{split}
	& \quad \tr \Bigl[ \CH^{-1} \, \bigl( \CH + \tilde{\CH} \bigr)^{-1} \, \nabla^{}_{\!(I} \CH \, \CH^{-1} \, \nabla^{}_{\!\!J)} \CH \Bigr] \\[2pt]
	& = \sum_{n=0}^\infty \tr \biggl\{ \Bigl[ \CH^{-1} \, \bigl( \mathbb{1} + \CH \bigr)^{-n-1} \Bigr] \, \nabla^{}_{\!(I|} \CH \, \Bigl[ \bigl( \mathbb{1} - \CH \bigr)^n \, \CH^{-1} \Bigr] \, \nabla^{}_{\!|J)} \CH \biggr\}\,,
\end{split}
\end{align}

In the equal-metric limit with $H^{}_{IJ} = G^{}_{IJ}$\,, we find
\be \label{eq:ghr}
	\beta^G_{IJ} = \beta^H_{IJ} = \frac{g^2}{4 \, \pi} \, R^{}_{IJ} + O(g^4)\,,
\ee
the beta-functional in Eq.~\eqref{eq:ghr} coincides with the beta-functional associated with the target-space metric field in two-dimensional relativistic sigma model. 
This is reminiscent of stochastic quantization of Euclidean field theories. In the limit $H \rightarrow G$\,, the bosonic part of the supersymmetric sigma model \eqref{eq:ssusyz2} is
\be \label{eq:sb}
	S^\text{B} = \frac{1}{2 \, g^2} \int dt \, d^2 \mathbf{x} \, \Bigl( \p_t  X^I \, \p_t X^J - \Box X^I \, \Box X^J \, \Bigr) \, G_{IJ}  \,.
\ee
 The potential term of the action \eqref{eq:sb} is a complete square of $\Box X^I$\,. Setting $\Box X^I = 0$ gives rise to the equation of motion from varying $X^I$ in the two-dimensional relativistic, Euclidean sigma model. This implies that the action \eqref{eq:sb} is at the detailed balance as in the theory of stochastic quantization. A direct consequence of the detailed balance is that the $(2+1)$-dimensional field theory at $z=2$ inherits the renormalization group flow of the associated two-dimensional Euclidean field theory. This explains why Eq.~\eqref{eq:ghr} takes the same form as the beta-functional in the two-dimensional relativistic sigma model. See also related discussions at the end of Section~\ref{sec:onnlsmz2}.

When the target-space is an $(N-1)$-sphere, we have
\be \label{eq:rgr}
	G^{}_{IJ} = \frac{1}{g^2} \lr \delta^{}_{IJ} + \frac{Y^I \, Y^J}{1 - Y \cdot Y} \rr,
		\qquad%
	R^{}_{IJ} = (N-2) \, g^2 \, G^{}_{IJ}\,.
\ee
Moreover, setting $H_{IJ} = \zeta \, G_{IJ}$\,, we find that the action \eqref{eq:se} reduces to the supersymmetric $O(N)$ NLSM action \eqref{eq:susyon}.
Plugging Eq.~\eqref{eq:rgr} and $H_{IJ} = \zeta \, G_{IJ}$ into Eq.~\eqref{eq:betaGH}, the beta-functions of $g$ and $\zeta$ in Eq.~\eqref{eq:betagsusy} are recovered.

\section{Quantum Critical Membranes} \label{sec:qcm}

In order to explore the possibility whether the three-dimensional $\CN=1$ sigma model \eqref{eq:ssusyz2} defined around a $z=2$ Lifshitz point can be promoted to describe quantum critical membranes, it is required that we couple the sigma model to worldvolume supergravity. In this section, we will present a few preliminary ingredients for constructing such a worldvolume gravity consistent with the $z=2$ Lifshitz scaling.

In this section, as an illustration of the concept, we couple the bosonic sector \eqref{eq:sbsusy} of the $\CN=1$ supersymmetric sigma model to worldvolume gravity. Due to the fundamental anisotropy between the worldvolume space and time, the appropriate worldvolume gravity has to be defined on a spacetime manifold foliated by codimension-one leaves. Such leaves are slices of constant time, and they are stacked along a preferred time direction. This type of gravity theories with a preferred time direction but \emph{no} local boost symmetry is referred to as Ho\v{r}ava gravity \cite{Horava:2008ih, Horava:2009uw}, which arises from gauging the foliation-preserving diffeomorphisms that act infinitesimally on the worldvolume coordinates $(t,\, \mathbf{x})$ as
\be \label{eq:fpd0}
	\delta t = \zeta (t)\,,
		\qquad%
	\delta x^i = \xi (t \,, \mathbf{x})\,.
\ee

We briefly review (non-projectable) Ho\v{r}ava gravity below.
The dynamics of Ho\v{r}ava gravity is encoded by the ADM variables, same as in the Hamiltonian formulation of general relativity. The ADM variables include the lapse function $N$, the shift vector $N^{}_i$\,, and the spatial metric $\gamma^{}_{ij}$\,, which transform under the foliation-preserving diffeomorphisms as
\begin{subequations}
\begin{align}
	\delta N & = \p^{}_i \bigl( \zeta \, N \bigr) + \xi^i \, \p^{}_i N\,, \\[2pt]
	\delta N^{}_i & = \p^{}_t \bigl( \zeta \, N^{}_i \bigr) + \xi^j \, \CD^{}_{\!j} N^{}_i + N^{}_j \, \CD^{}_i \xi^j + \gamma^{}_{ij} \, \p^{}_t \xi^j\,, \\[2pt]
	\delta \gamma^{}_{ij} & = \zeta \, \p_t \gamma^{}_{ij} + \CD^{}_i \, \xi^{}_j + \CD^{}_{\!j} \, \xi^{}_i\,,
\end{align} 
\end{subequations}
where $\CD^{}_i$ is the spatial covariant derivative satisfying $\CD^{}_i \gamma^{}_{jk} = 0$\,. At a $z=2$ Lifshitz point, we assign the following engineering scaling dimensions to the ADM variables:
\be
	[N] = 0\,,
		\qquad%
	[N^{}_i] = \frac{1}{2}\,,
		\qquad%
	[\gamma^{}_{ij}] = 0\,.
\ee
Recall that we have chosen $[t] = -1$ and $[\mathbf{x}] = -1/2$ at the $z=2$ fixed point, such that energy has a unit scaling dimension. 
In addition to the Lifshitz scaling symmetry, we also impose the time-reversal symmetry,
\be
	t \rightarrow -t\,,
		\qquad%
	N \rightarrow N\,,
		\qquad%
	N^{}_i \rightarrow - N^{}_i\,,
		\qquad%
	\gamma^{}_{ij} \rightarrow \gamma^{}_{ij}\,,
\ee
together with the local anisotropic Weyl scaling symmetry,
\be
	N \rightarrow e^{2 \, \omega (t, \mathbf{x})} \, N\,,
		\qquad%
	N^{}_i \rightarrow e^{2 \, \omega(t, \mathbf{x})} \, N^{}_i\,,
		\qquad%
	\gamma^{}_{ij} \rightarrow e^{2 \, \omega(t, \mathbf{x})} \, \gamma^{}_{ij}\,.
\ee
The most general worldvolume gravitational action consistent with the above symmetries is
\be \label{eq:swg}
	S_\text{gravity} = \frac{1}{2 \, \CG} \int dt \, d^2 x \, N \sqrt{\gamma} \ls \lr K^{ij} \, K^{}_{ij} - \frac{1}{2} \, K^2 \rr + \alpha \lr R + \nabla^i a^{}_i \rr^{\!2} \, \rs,
\ee
where there are two marginal couplings $\CG$ and $\alpha$\,. 
Here, $R$ is the Ricci scalar associated with the spatial metric $\gamma^{}_{ij}$\,, $K^{}_{ij}$ is the extrinsic curvature with
\be
	K^{}_{ij} = \frac{1}{2 \, N} \bigl( \p_t \gamma^{}_{ij} - \CD^{}_i N^{}_j - \CD^{}_j N^{}_i \bigr)\,,
		\qquad%
	K = \gamma^{ij} \, K^{}_{ij}\,,
\ee
and $a_i$ is the acceleration with
\be
	a^{}_i = - \p^{}_i \ln N\,.
\ee
In addition to the terms in Eq.~\eqref{eq:swg}, there are also five independent boundary terms that are Weyl invariant \cite{Griffin:2011xs, Baggio:2011ha}, which may have non-trivial contribution to the candidate membrane theory when coupled to dilaton-like background fields. 

We now couple the matter field to the worldvolume geometry by covariantizing the matter action \eqref{eq:ssusyz2} with respect to the worldvolume foliation-preserving diffeomorphisms in \eqref{eq:fpd0}. The resulting action is
\be \label{eq:smatter}
	S_\text{matter} = \frac{1}{2 \, g^2} \int dt \, d^2 \mathbf{x} \, N \sqrt{\gamma} \, \Bigl( d_t  X^\mu \, d_t X^\nu \, G_{\mu\nu} - \Box X^\mu \, \Box X^\nu \, H_{\mu\rho} \, G^{\rho\sigma} \, H_{\sigma\nu} \Bigr) \,,
\ee
where
\be
	d^{}_t X^\mu = \frac{1}{N} \bigl( \p_t - N^i \, \p^{}_i \bigr) X^\mu\,,
		\qquad%
	\Box X^\mu = \CD^i \p^{}_i X^\mu + \Theta^\mu{}_{\rho\sigma} \, \p^{}_i X^\rho \, \p^{}_i X^\sigma\,.
\ee
Here, we have introduced the Minkowskian index $\mu = 0\,, \cdots, \, d-1$ and the target-space gains a spacetime interpretation, with $X^\mu$ playing the role of the embedding coordinates that map the worldvolume to the target space. Coupling the matter sector described by Eq.~\eqref{eq:smatter} to the worldvolume gravity described by Eq.~\eqref{eq:swg} defines the sigma model for the bosonic sector of the quantum critical membrane. In general, the couplings in the gravity sector, including $\CG$ and $\alpha$\,, and possibly also the couplings in front of the boundary terms, will be promoted to be dilatational fields.

The complete sigma model describing the quantum critical supermembrane requires us to formulate a worldvolume supergravity whose bosonic sector can be identified with the action \eqref{eq:swg}. Some of the basic ingredients of supersymmetrizing Ho\v{r}ava gravity have already been constructed in \cite{Frenkel:2020dic}, in a different context of cohomological quantum gravity and Hamilton-Perelman's Ricci flow equation. 
The quantization of the worldvolume (super)gravity can be implemented using the new heat kernel method in \cite{Grosvenor:2021zvq}, which is designed for Lifshitz-type quantum field theories on a curved, foliated manifold. This heat kernel method generalizes the previous work on relativistic QFTs that uses a Fourier transform covariantized with respect to curved geometries \cite{Gusynin:1989ky}. Such a covariant notion of the Fourier transform is first introduced in the studies of pseudo-differential operators in functional analysis \cite{widom1978families, widom1980complete}. In \cite{Grosvenor:2021zvq}, a Fourier transform covariantized with respect to a foliated geometry is constructed, essentially by replacing the phase factor in the exponential of Eq.~\eqref{eq:osigmaeqn} with the associated phase function adapted to the foliation structure, which leads to a covariant heat kernel method suitable for quantizing Lifshitz sigma models describing the quantum critical membrane. 

One challenge of implementing the heat kernel method to evaluate the one-loop effective action for the bosonic version of Ho\v{r}ava gravity comes from the fact that higher-order Seeley-Gilkey coefficients are needed, in comparison to the analogous calculation in relativistic QFTs. For example, the beta-function calculation of the three-dimensional Ho\v{r}ava gravity described by Eq.~\eqref{eq:swg} requires evaluating the Seeley-Gilkey coefficient $E_4$\,, instead of $E_2$ for three-dimensional relativistic QFTs. The calculation capacity required for evaluating the higher-order Seely-Gilkey coefficients increases rapidly. However, in the superspace formalism, as we have seen for the renormalization of the matter sector in Section~\ref{sec:sgcea}, it is still $E_2$ that we need to compute, which makes it much more manageable to derive the complete one-loop effective action (including the gravitational sector) for the quantum critical supermembrane.\,\footnote{Note that encouraging progress has been made for quantizing bosonic Ho\v{r}ava gravity. For the quantization of projectable Ho\v{r}ava gravity in three dimensions, see, \emph{e.g.}, \cite{Benedetti:2013pya, Barvinsky:2015kil, Barvinsky:2017kob, Griffin:2017wvh}. The quantization of Ho\v{r}ava gravity becomes even more intricate in four dimensions, even for the projectable case \cite{Barvinsky:2019rwn}. If one could construct a superspace formalism for four-dimensional Ho\v{r}ava gravity, it would also be much more manageable to apply the heat kernel method in \cite{Grosvenor:2021zvq} to evaluate the associated RG structure, as one would only need to compute $E_4$ instead of $E_6$ for the bosonic case. Moreover, the presence of supersymmetry may also provide an opportunity for us to understand the low-energy (possibly strongly-coupled) behavior of Ho\v{r}ava gravity and its potential of flowing toward General Relativity.}

\section{Conclusions} \label{sec:concl}

In this paper, we studied the renormalization of three-dimensional $\CN = 1$ supersymmetric nonlinear sigma model that exhibits a worldvolume Lifshitz scaling, with the dynamical critical exponent $z=2$\,. The sigma model is coupled to a target-space geometry equipped with a bimetric structure. The beta-functionals of the bimetric fields are evaluated in Eq.~\eqref{eq:betaGH} using a covariant heat kernel method. We also commented on coupling the bosonic sector of the Lifshitz sigma model to the worldvolume Ho\v{r}ava gravity. 

It is of immenient interest to supersymmetrize the worldvolume Ho\v{r}ava gravity discussed in Section \ref{sec:qcm} and consider its coupling to the supersymmetric sigma model at quantum criticality. This would provide a framework that describes the quantum critical supermembrane as a candidate high-energy completion of the relativistic membrane theory. It is important to understand the BRST symmetry of the quantum critical supermembrane and analyze the physical spectrum of the theory in flat target spacetime. This will allow us to determine the conditions under which the theory is unitary,\,\footnote{A preliminary analysis of the BRST symmetry of the two-dimensional Lifshitz sigma model for strings in a bimetric spacetime \cite{Yan:2021xnp} does not seem to result in a unitarity theory: on the two-dimensional worldsheet, the foliation-preserving diffeomorphisms do not appear to be sufficient to remove the unphysical degrees of freedom due to the ``wrong" sign of the temporal component of the embedding coordinate $X^\mu$\,. There is a possible way out of this obstruction: instead of coupling the two-dimensional sigma model to topological non-projectable Ho\v{r}ava gravity, one may attempt to couple the sigma model to Ho\v{r}ava gravity with an extended $U(1)$ gauge symmetry \cite{Horava:2010zj}. Such a worldsheet gravity with "nonrelativistic" general covariance, whose associated diffeomorphisms, upon gauge fixing, might be sufficient to eliminate the unphysical degrees of freedom. Further analysis is required to determine whether this intuition applies to the quantum critical membrane.} which is important for us to understand the possibility of defining a self-consistent bimetric quantum gravity in spacetime. On the other hand, it is also possible that such physical conditions might require the metric fields to be equal to each other, in which case the bosonic part of the sigma model satisfies the detailed balance condition and the ground state of the quantum critical membrane reproduces the partition function of bosonic string theory, in reminiscence of the stochastic quantization of Euclidean QFTs \cite{Horava:2008ih}. 

Another important question is whether the quantum critical supermembrane reproduces the relativistic supermembrane theory in the deep IR, which is strongly coupled. It is therefore expected that nonperturbative approaches are required for establishing such a mechanism. It would be enlightening to explore how supersymmetry may help restore the relativistic general covariance in low energies.
Note that this requires understanding RG flows away from the Weyl fixed points in string/M-theory. In string theory, it is important that the worldsheet theory is conformally invariant, and the vanishing beta-functionals correspond to classical equations of motion governing the background field dynamics in the target space. This construction leads to the first-quantized form of string theory. Even though the sum over Riemann surfaces already captures all possible interactions, this first-quantized formulation of string theory may not capture many of the nonperturbative effects. String field theory as a path integral over all configurations of spacetime background fields is then proposed as a generalized framework, which is elegantly realized by constructing the path integral for a spacetime action whose equation of motion reproduces the BRST-invariant form of the theory \cite{Witten:1992qy}. Conceptually, the RG of the worldsheet QFT encodes transitions between different string theory vacua, which arise as RG fixed points. In principle, the same considerations also apply to the theory of quantum critical membranes.

In general, it is rather difficult to gain a thorough understanding of the spacetime physics corresponding to the worldvolume RG. An essential first step toward the construction of the RG flows is to map out available critical points connected by these RG flows. It is possible that the Lifshitz-type sigma model describing membranes at quantum criticality defines such a perturbative RG fixed point, and the relevant deformation controlled by a functional version of the worldvolume speed of light generates an RG flow toward the relativistic fixed point, before flowing toward different string theories. It would be intriguing to understand how this potential RG flow can be embedded in the framework of second-quantized strings/membranes. Moreover, it would also be fascinating to consider other more exotic nonrelativistic RG fixed points defined by higher-dimensional sigma models at quantum criticality, which may lead to candidate worldvolume theories that UV complete various relativistic $p$-brane theories and novel target-space geometries.  

\acknowledgments

I would like to thank Petr Ho\v{r}ava and Charles Melby-Thompson for collaborations and many inspirations at an early stage of this project. It is also a pleasure to thank Eric Bergshoeff, Kevin Grosvenor, Niels Obers, Gerben Oling, Benjamin Ponedel, Stephen Randall, and Aaron Szasz for useful discussions. Special thanks to Kevin Grosvenor for discussions and collaborations on the heat kernel method. This work has been presented at the workshop ``\emph{What is new in gravity?}" hosted by the Niels Bohr Institution in Helsing{\o}r, Denmark. The author would like to thank all the participants at the workshop for stimulating discussions. This work is supported by the European Union's Horizon 2020 research and innovation programme under the Marie Sk{\l}odowska-Curie grant agreement No 31003710. Nordita is supported in part by NordForsk.

\newpage

\bibliographystyle{JHEP}
\bibliography{rslsm}

\end{document}